\documentclass[twocolumn,           
               showpacs,            
               preprintnumbers,     
               aps,                 
               prd,          	    
               letterpaper,             
               groupaddress,      
               nofootinbib,         
               tightenlines,        
               floats,floatfix      
               ]{revtex4-1}

\usepackage{amsmath,amssymb}
\usepackage{latexsym}
\usepackage{graphicx}
\usepackage{epstopdf}
\usepackage{graphicx}
\usepackage{dcolumn}
\usepackage{bm}
\usepackage{subfigure}
\usepackage{enumerate}
\usepackage[draft=false]{hyperref}

\begin{document}


\title{$\Phi^{4}$ Oscillatons}

\author{Susana Valdez-Alvarado}
 \email{svaldez@fisica.ugto.mx}
\author{L. Arturo Ure\~na-L\'opez}%
 \email{lurena@fisica.ugto.mx}
\affiliation{Departamento de F\'{\i}sica, DCI, Campus Le\'on,
  Universidad de Guanajuato, 37150, Le\'on, Guanajuato,
  M\'exico}%
\author{Ricardo Becerril}
 \email{becerril@ifm.umich.mx}
\affiliation{Instituto de F\'{\i}sica y Matem\'{a}ticas,
        Universidad Michoacana de San Nicol\'as de Hidalgo. Edificio C-3,
        Cd. Universitaria, 58040, Morelia, Michoac\'{a}n, M\'exico}%

\date{\today}

\begin{abstract}
We solve numerically the Einstein-Klein-Gordon system with spherical
symmetry, for a massive real scalar field endowed with a
quartic self-interaction potential, and obtain the so-called
$\Phi^4$-oscillatons which is the short name for oscillating soliton
stars. We analyze numerically the stability of such oscillatons, and
study the influence of the quartic potential on the  behavior of both,
the stable (S-oscillatons) and  unstable (U-oscillatons) cases under
small and strong radial perturbations. 
\end{abstract}

\pacs{04.40.-b,04.25.D-,95.30.Sf,95.35.+d}
\maketitle

\section{\label{sec:introduction-1}Introduction}
\emph{Oscillatons} are non-singular and asymptotically flat solutions
of the Einstein-Klein-Gordon (EKG) equations, in which both metric and
scalar field are fully
time-dependent\cite{Seidel:1991zh,UrenaLopez:2002gx,Alcubierre:2003sx,
  UrenaLopez:2001tw,Balakrishna:2007mr}. They   
can be considered as gravitationally bounded objects made of a
real scalar field in the classical regime, and were first discovered
by Seidel and Suen\cite{Seidel:1991zh}. Oscillatons should be
distinguished from their complex counterparts, the so-called boson
stars, for which the space-time geometry remains
static\cite{Seidel:1993zk,Balakrishna:2006ru,Colpi:1986ye,Guzman:2004jw}. In
the literature we can find other works describing these bound
objects and their stability properties\cite{Grandclement:2011wz,Fodor:2010hg,Fodor:2009kg,Masso:2005zg,Obregon:2004ty,Matos:2008ag,
  Page:2003rd}. Also
there have been many works about numerical evolution of scalar fiels in
different cosmological and astrophysical
context\cite{Guzman:2009zz,Bernal:2006ci,Bernal:2006it}.

Ref.~\cite{Alcubierre:2003sx} presented an exhaustive study of
oscillatons for the simplest case of a massive, non-self interacting,
scalar field. It is shown in there that, as in the case of boson
stars, oscillatons can be classified into stable (S-branch) and
unstable (U-branch) configurations. S-oscillatons are stable
configurations under small radial perturbations, and they typically
migrate to other S-profiles if strongly perturbed. On the other hand,
U-oscillatons are intrinsically unstable: they migrate to the S-branch
if their mass is moderate, but they may collapse into black holes if
their mass is large enough. 

The case of oscillatons with a self-interaction term in the
scalar field potential remains without study. This is in contrast to
the case of boson stars, whose properties are well known even in the
self-interacting
case\cite{Colpi:1986ye,Balakrishna:1997ej,Guzman:2004jw,Balakrishna:1997ej}. It
is the main purpose of this paper to cover this omission, and to study the
properties of self-interacting oscillatons. For that we make use of
techniques and methods that parallel those used in boson star studies
and for non-interacting
oscillatons\cite{Seidel:1990jh,Balakrishna:1997ej,Alcubierre:2003sx}, 
which will ease the comparison between the two types of scalar
objects.

A summary of the paper is as follows. In Sec.~\ref{sec:math-backgr},
we present the equations motion of the self-interacing scalar
field. In Sec.~\ref{sec:equil-conf} we show the resuls of the
equilibrium configuration and their representative properties, for
different values of the self-interaction parameter. In
Sec.~\ref{sec:numer-evol-self} we present the results of the dynamical
evolution of oscillatons, and the separate analysis of (stable)
S-oscillatons and (unstable) U-oscillatons. Finally,
Sec.~\ref{sec:conclusions} is devoted to overall conclusions.

\section{\label{sec:math-backgr}Mathematical Background}
The action that describes our self-gravitating system is
\begin{equation}
  I = \int d^{4}x \sqrt{-g} \left( \frac{R}{16\pi G} - \frac{1}{2}
    g^{\alpha\beta} \Phi_{,\alpha} \Phi_{,\beta} + V(\Phi) \right) \,
  , \label{axion}
\end{equation}
where $V(\Phi)$ is the scalar potential,
\begin{equation}
  V(\Phi) = \frac{1}{2} m^{2}_{\Phi} \Phi^{2} + \frac{1}{4} \lambda
  \frac{\Phi^{4}}{4} \, , \label{potential}
\end{equation}
$m_{\Phi}$ denotes the mass of the scalar field, and $\lambda$ is the
quartic interaction parameter. We shall be interested in the
spherically symmetric case for which the metric is written as
\begin{equation}
  ds^{2} = -\alpha^2(t,x) dt^{2} + a^2(t,x) dr^{2} + r^{2} ( d\theta^{2} +
  \sin^{2} \theta \, d\psi^{2} ) \, , \label{metric}
\end{equation}
where $a(t,r)$ is the radial function, and $\alpha(t,r)$ is the lapse
function.

\subsection{\label{sec:evolution-equations}Evolution Equations}
The energy-momentum tensor for the scalar field
$\Phi(t,r)$ endowed with a scalar field potential $V(\Phi)$ is defined
as
\begin{eqnarray}
  T_{\mu \nu} = \Phi_{,\mu} \Phi_{,\nu} - \frac{1}{2}
  g_{\mu \nu} \left[ \Phi^{,\sigma} \Phi_{,\sigma} + m^{2}_{\Phi}
  \Phi^{2} + \frac{1}{2} \lambda \Phi^{4} \right] \, .
\end{eqnarray}
We can identify the Klein-Gordon (KG) equation with the conservation
equations of the scalar field  energy-momentum tensor,
\begin{equation}
  {T^{\mu \nu}}_{;\nu} = \Phi^{,\mu} \left( \Box \Phi -
    \frac{dV}{d\Phi} \right) = 0 \, , \label{conservation}
\end{equation}
where $\Box = (1/\sqrt{-g}) \partial_\mu ( \sqrt{-g} \partial^\mu)$ is
the d'Alembertian operator. Following Ref.~\cite{Alcubierre:2003sx},
we introduce the first order variables
\begin{equation}
\Psi=\Phi^\prime \, , \quad \Pi=a \dot{\Phi} /\alpha \, , \label{nwe}
\end{equation}
in order to write appropriate evolution equations for the scalar
fields. Hereafter, a prime denotes derivative with respect to $r$, and
a dot denotes derivative with respect to $t$. The KG equation is then
represented by the following set of first order differential equations 
\begin{subequations}
\label{eq:set}
\begin{eqnarray}
  \dot{\Phi} &=& \frac{\alpha}{a}\Pi \, , \label{set1}\\
  \dot{\Pi} &=& \frac{1}{x^2} \left( \frac{x^2 \alpha \Psi}{a}
  \right)^\prime - a\alpha(\Phi + \Lambda \Phi^3 ) \, , \label{set2} \\
  \dot{\Psi} &=& \left(\frac{\alpha\Pi}{a} \right)^\prime \,
  . \label{set3}
\end{eqnarray} 
\end{subequations}

As for the metric functions, we will use the so-called Hamiltonian
constraint for the radial function,
\begin{equation}
  \frac{a_{,x}}{a} = \frac{1-a^2}{2x} + \frac{x}{4} \left[ \Pi^2 + \Psi^2
      +a^2 \left( \Phi^2 + \frac{1}{2} \Lambda \Phi^4 \right) \right]
    \, , \label{lapso}
\end{equation}
and the  polar-areal slicing condition for the lapse function,
\begin{equation}
  \frac{\alpha_{,x}}{\alpha} = \frac{a^2-1}{x} - a^2 x \left(
      \frac{\Phi^2}{2} + \frac{1}{4} \Lambda \Phi^4 \right) +
    \frac{a_{,x}}{a} \, . \label{shift}
\end{equation}
For numerical purposes, in all the above equations, we have introduced the
dimensionless quantities $r=x/m_{\Phi}$, $t\rightarrow t/m_{\Phi}$,
$\Phi\rightarrow \Phi/\sqrt{\kappa_0}$, and $\Lambda \equiv
\lambda/(8\pi Gm^2_{\Phi})$. Eqs.~(\ref{eq:set}-\ref{shift}) are the
total set of evolution equations we shall use to explore the
properties of self-interacting oscillatons.

\section{\label{sec:equil-conf}Equilibrium Configuration}
Before we evolve the EKG equations, we turn our attention to
equilibrium configurations of the self-interacting oscillatons. For
that, we consider momentarily a different set of variables: $A(t,x) =
a^2$, and $C(t,x) = a^2/\alpha^2$. In terms of these variables,
the Einstein equations, $G_{\alpha \beta} = 8\pi T_{\mu \nu}$, read
\begin{subequations}
\label{eq:einstein}
\begin{eqnarray}
  \dot{A} &=& x A \dot{\Phi} \Phi^\prime \, , \label{einstein1} \\
  A^\prime &=& \frac{xA}{2} \left[ C\dot{\Phi}^{2} + {\Phi^\prime}^2 +
    A \left( \Phi^2 + \frac{1}{2} \Lambda \Phi^4 \right) \right] -
  \frac{A}{x}(1-A) \, , \label{einstein2} \\
  C^\prime &=& \frac{2C}{x} \left[ 1 + A \left( x^2 \left[
        \frac{1}{2} \Phi^2 + \frac{1}{4} \Lambda \Phi^4 \right] - 1
    \right) \right] \ , \label{einstein3}
\end{eqnarray}
\end{subequations}
whereas the KG is the single equation
\begin{eqnarray}
  \Phi^{\prime \prime} &=& C \ddot{\Phi} + \dot{C}\dot{\Phi} -
  \frac{\Phi^\prime}{x} \left[ 1 + A \left( 1 - x^2 \left[ \frac{1}{2}
        \Phi^2 + \Lambda \frac{1}{4} \Phi^4 \right] \right) \right]
  \nonumber \\
& & + A ( \Phi + \Lambda \Phi^3 ) \label{kleingordon} \, . 
\end{eqnarray}

The above equations will be solved by introducing in
Eqs.~(\ref{eq:einstein}) and~(\ref{kleingordon}), the following
Fourier expansions for the scalar field and the metric functions,
\begin{subequations}
\label{eq:fourier}
\begin{eqnarray}
  \Phi(t,x) &=& \sum_{j=1}^{J_{max}} \phi_{2j-1}(x) \cos((2j-1)\omega t) \\ 
  A(t,x) &=& \sum_{j=0}^{J_{max}} A_{j}(x) \cos(2j\omega t) \\ 
  C(t,x) &=& \sum_{j=0}^{J_{max}} C_{j}(x) \cos(2j\omega t) 
\end{eqnarray}
\end{subequations}
where $\omega$ is called the fundamental frequency, and $J_{max}$ is
the mode at which the Fourier series are truncated. We define $C = C
\Omega^{-2}$, with $\Omega=\omega/m_{\Phi}$. 

After some lengthy but straightforward algebra, we obtain first-order
differential equations for the coefficients $A_{0}$, $C_{2j}$,
$\Phi_{2j-1}$ and $\Psi_{2j-1}$, where
$\Psi_{2j-1}=\Phi^\prime_{2j-1}$. This system of equations can be
solve with the aid of an integrator, like {\it odeint}, that uses an
adaptive step-size control which allows us to achieve some predetermined
accuracy in the solution with minimum computational effort
(see~\cite{numerical}). We also use a second order {\it Runge-Kutta} method
(RK2) in order to verify the convergence of our code, since this
method uses a fix step-size. Once the profiles of $A_{0}$ and
$\Phi_{2j-1}$ are known, we will get the subsequent values of
$A_{2j}$, for $j\neq 0$, from the algebraic
restriction~(\ref{einstein1}).

\subsection{\label{sec:boundary-conditions}Boundary Conditions}

Appropriate boundary conditions can be obtained by analyzing the
regularity of the functions $A(t,x)$, $C(y,x)$, and $\Phi(t,x)$ at
$x\rightarrow 0$. First, we realize that $A(t,x=0)=1$, which
implies $A_{0}(x=0)=1$, and $A_{j}(x=0)=0$ for  $j\geq1$. Second, we
find that $\Phi^\prime (t,x=0)=0$, so that $\phi^\prime_{j}(x=0)=0$
$\forall j$. 

From asymptotic flatness, we know that $A(t,x\rightarrow\infty)=1$,
which leads to $A_{0}(x \rightarrow \infty) =1$, and
$A_{j} (x \rightarrow \infty) = 0$ for $j\geq 1$. The scalar field
must vanish when $x\rightarrow\infty$, and this implies
$\phi_{j}(x\rightarrow\infty)=0$ $\forall j$. Likewise, we expect
$C(t,x \rightarrow \infty) = \Omega^{2}$, and then $C_{0}(x
\rightarrow \infty) =\Omega^{2}$, whereas $C_{j}(x \rightarrow
\infty)=0$ for $j\geq 1$. Actually, the fundamental frequency $\Omega$
is always an output value obtained after the solution of
Eqs.~(\ref{einstein2}),~(\ref{einstein3}), and~\ref{kleingordon}).

Note that there are not boundary conditions for $C_{j}(x)$ and
$\phi_{j}(x)$ at $x=0$. We need to find these values in such manner
that all boundary conditions at $x\rightarrow\infty$ are
satisfied. For each solution, we will fix the value of the first
scalar Fourier coefficient $\phi_{1}(0)$, and adjust the rest of the
coefficients $\phi_{j\leq 2}$ and $C_{j\geq 0}$ until we satisfy the
boundary conditions\cite{UrenaLopez:2002gx}; we use a non-linear
shooting method to find these values. Thus, we obtain a set of
eigenvalues for each value $\phi_{1}(0)$, which in turn defines
uniquely a given configuration.

\subsection{\label{sec:numerical-results}Numerical Results}

We truncate the Fourier expansions~(\ref{eq:fourier}) at
$J_{max}=2$. Then, according to what we said about boundary
conditions, we need to calculate the value of the coefficients
$C_{0}$, $C_{2}$, $C_{4}$, and $\phi_{3}$, for each value of the
$\Lambda$ parameter. Table~\ref{table} contains the values obtained
once the boundary conditions were satisfied at $x_{max}=30$ for the
fixed value $\phi_{1}(0)=0$.$28$. 

\begin{table}[htp]
\begin{tabular}{c|cccc}\hline
& $\Lambda=$0.0 & $\Lambda=$1.0 & $\Lambda=$2.0 & $\Lambda=$3.0 \\ 
\hline
$C_{0}(0)$ & 8.98e-01 & 8.91e-01 & 8.88e-01 & 8.84e-01 \\ 
$C_{2}(0)$ & 6.90e-05 & 1.16e-05 & 1.36e-05 & 3.54e-05 \\ 
$C_{4}(0)$ & 2.66e-06 & 7.32e-06 & 9.11e-06 & 1.68e-05\\ 
$\phi_{3}(0)$ & 5.77e-07 & 3.69e-06 & 6.85e-06 & 6.57e-07\\  \\
\end{tabular}
\caption{\label{table}Resulting values of the Fourier coefficients $C_{0}$,
  $C_{2}$, $C_{4}$, and $\phi_{3}$, see Eqs.~(\ref{eq:fourier}), after
  the integration of the equations of motion~(\ref{eq:einstein}); in
  all cases, $\phi_{1}(0)=0.28$, and $x_{max} = 30$. For completeness,
  the first column shows the usual, non-interacting, massive case; it
  can be verified that the values obtained coincide with those
  reported in\cite{Alcubierre:2003sx}.
}
\end{table}

\begin{figure}[htp]
  \includegraphics[width=0.5\textwidth]{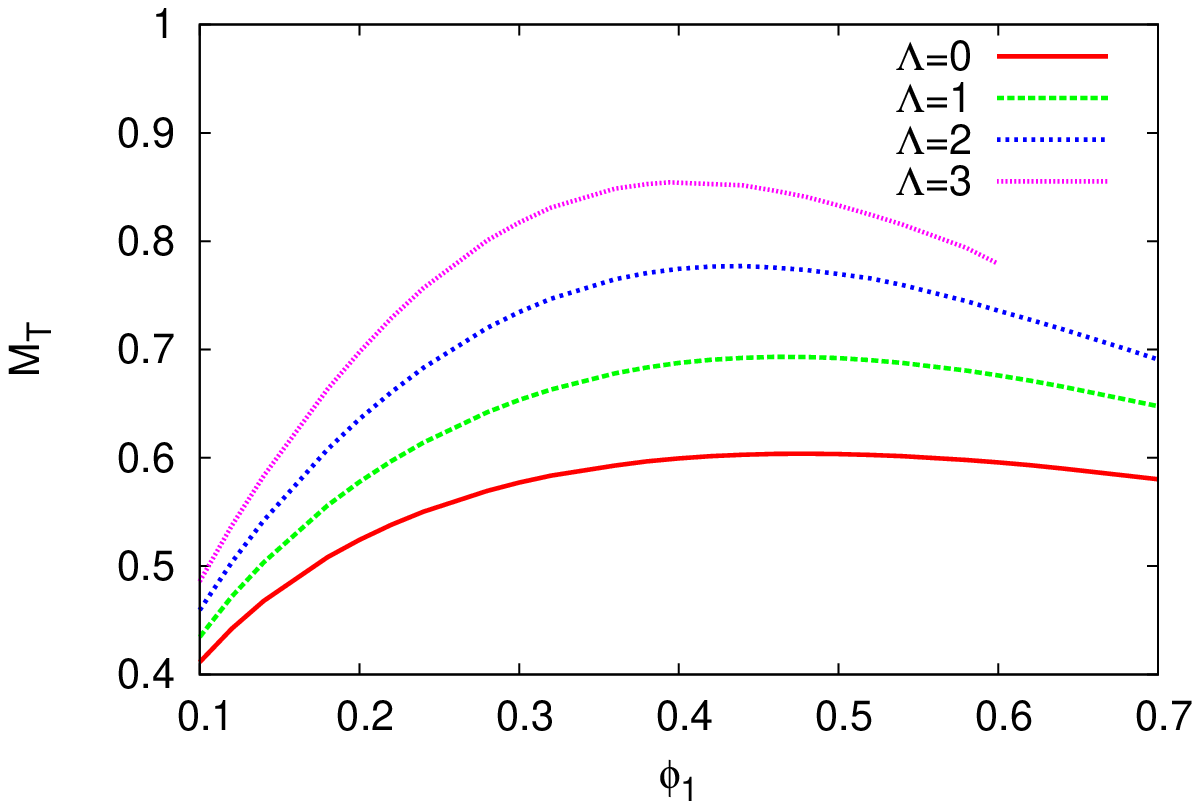}
  \includegraphics[width=0.5\textwidth]{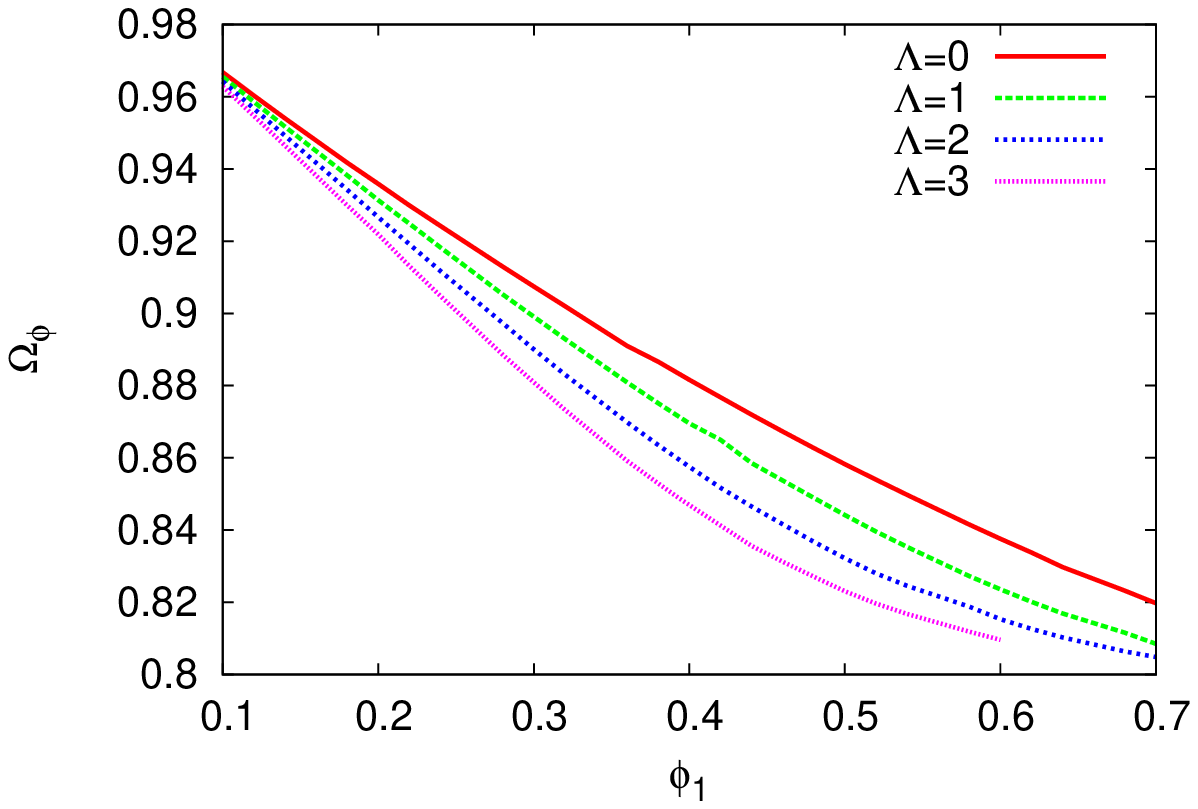}
  \caption{\label{masas} (Top) $M_{\Phi}$ and (Bottom) $\Omega_\phi$ as
    functions of the central value of the Fourier coefficient
    $\phi_{1}(0)$ for different values of $\Lambda$. We can observe in
    that there is a shift to the left of the critical value $\phi_{1
      c}(0)$, which corresponds to that equilibrium configuration with
    a critical mass $M_{\Phi c}$, as $\Lambda$ increases.} 
\end{figure}

In contrast to the non-interacting case, the increasing of $\Lambda$
complicates the solution of the equations of motion as we consider
larger values of $\phi_{1}$. As a consequence, in Fig.~\ref{masas} we
observe that, for $\Lambda=2,3$, we were able to compute the values of
masses and frequencies for scalar field values up to
$\phi_{1}(0)=0.6$, whereas for $\Lambda=0,1$ we could compute up to
$\phi_{1}=0.7$. In the same way, as the values of $\Lambda$ and of
$\phi_{1}(0)$ are increased, we must decrease the size of the numerical
domain, represented by $x_ {max}$.

\begin{figure}[!htp]
  \includegraphics[width=0.5\textwidth]{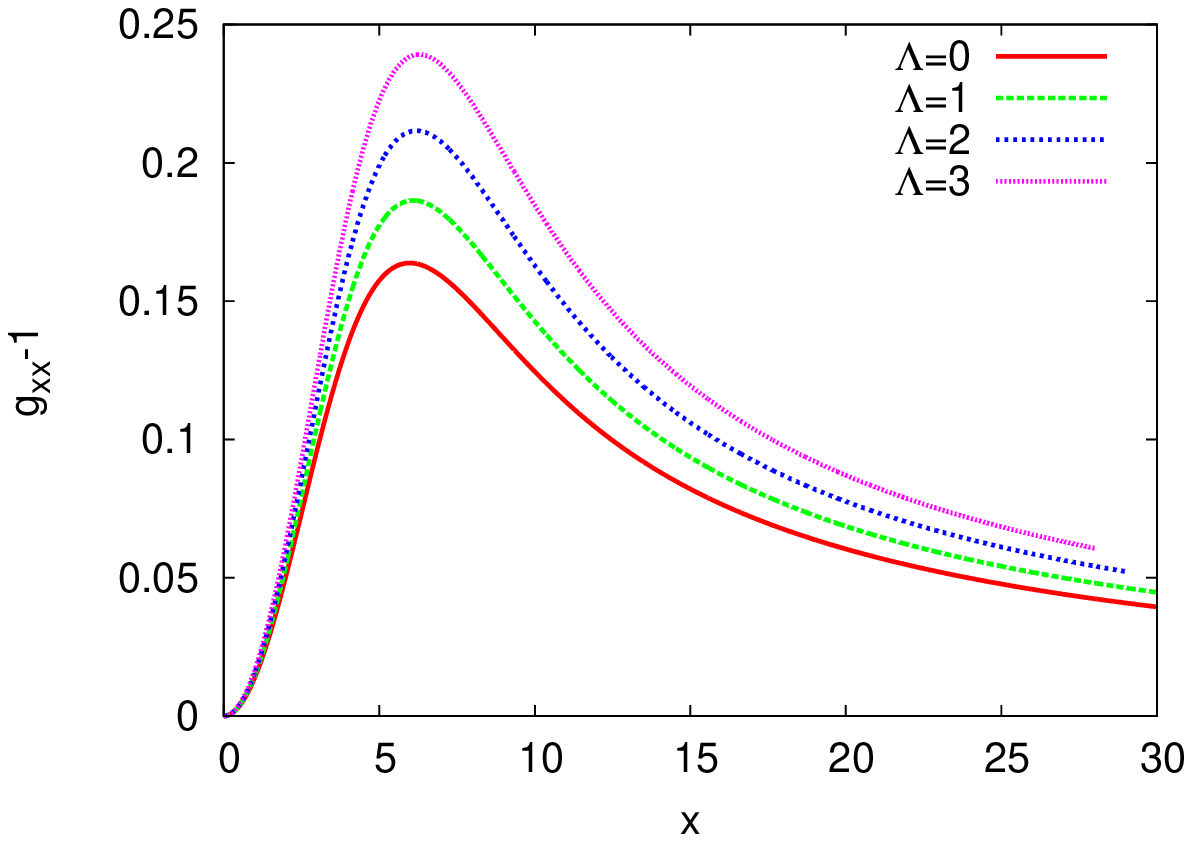}
  \includegraphics[width=0.5\textwidth]{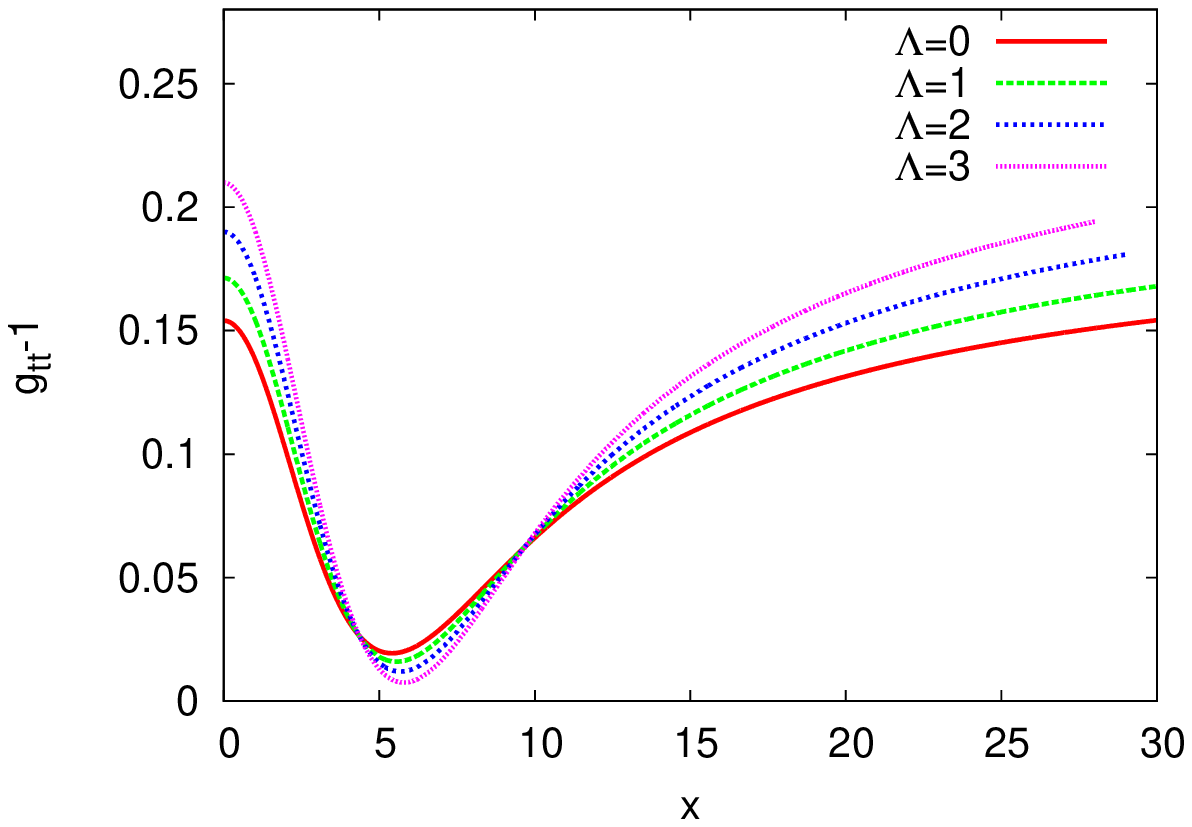}
  \includegraphics[width=0.5\textwidth]{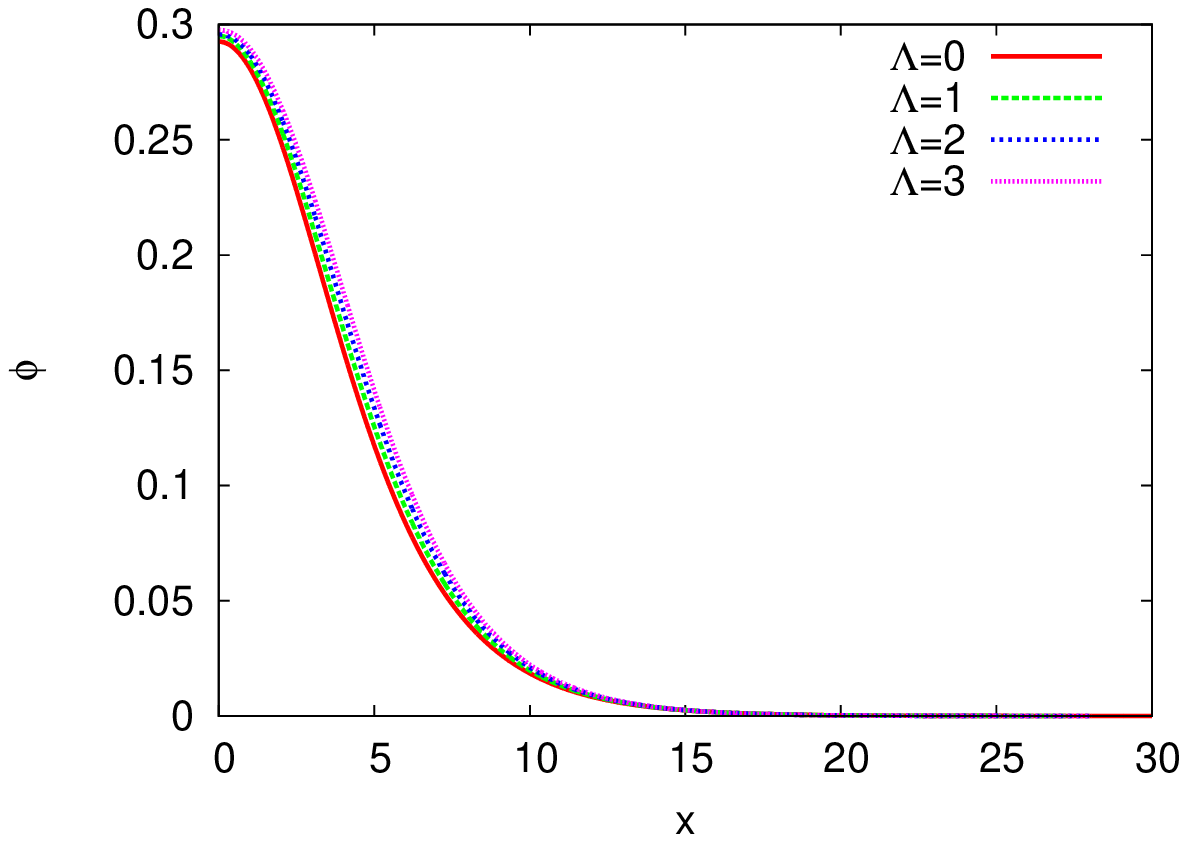}
  \caption{Metric functions (Top) $g_{rr}$, (Middle) $g_{tt}$, and
    (Bottom) the scalar field $\Phi$ of equilibrium configurations
    corresponding to different values of the quartic parameter
    $\Lambda$. For all examples, we set $\phi = 0.2$, $x_{max}=30$.}
    \label{funciones}
\end{figure}

We present in Fig.~\ref{funciones} the results for the metric
functions $g_{rr}$, $g_{tt}$, and the scalar field $\Phi$, for
configurations with $\phi_{1}(0)=0.2$ and $t=0$. This functions were
calculated for different values of the parameter $\Lambda$. Note that
the metric functions are non-singular at the origin and asymptotically
flat for large $x$.

Following \cite{UrenaLopez:2002gx},  the total mass $M_{T}$ is
calculated by the so-called Schwarzschild mass,
\begin{equation}
  M_{\Phi} = \frac{m^{2}_{Pl}}{m_{\Phi}} \lim_{x \rightarrow \infty}
  \frac{x}{2} \left[ 1 -A^{-1}(t,x) \right] \, , \label{eq:mass}
\end{equation}
and the fundamental frequency $\Omega$ is obtained from
\begin{equation}
 \Omega_{\Phi} = \lim_{x \rightarrow \infty}
  \frac{\sqrt{C(t,x)}}{A(t,x)} \, .
\end{equation}
In Fig.~\ref{masas}, we show the values obtained for the total masses
and fundamental frequencies as functions of the central value of the
first scalar field (Fourier) coefficient $\phi_{1}(0)$ and for
different values of $\Lambda$. We can see that the mass of a given
configuration increases for larger values of $\Lambda$ increases,
whereas the fundamental frequency $\Omega_{\Phi}$ decreases. 

In Table~\ref{masafrec}, we show the values of the fundamental
frequency $\Omega_\Phi$ and of the critical masses $M_{\Phi c}$ for
each value of $\Lambda$. The critical mass is defined as the maximum
mass attained by an equilibrium configuration; this maximum can be
seen in Fig.~\ref{masas} for each value of $\Lambda$. 

\begin{table}[htp]
\begin{tabular}{c|ccc}\hline
$\Lambda$ & $M_{\Phi c}$($m^{2}_{Pl}/m_{\Phi}$) & $\phi_{1 c}(0)$ &
  $\Omega_{\Phi}$ \\  
\hline
0.0 & 0.605 & 0.48 & 0.862\\ 
1.0 & 0.694 & 0.47 & 0.851\\ 
2.0 & 0.770 & 0.44 & 0.850\\ 
3.0 & 0.854 & 0.39 & 0.848\\
\end{tabular}
\caption{\label{masafrec} Resulting values of the critical mass
  $M_{\Phi c}$ for different $\Lambda$'s. As expected, the mass of the
  equilibrium configurations increase for larger values of $\Lambda$,
  whereas the fundamental frequency decreases. There is also a shift
  to lower values of the critical value $\phi_{1 c}(0)$ as $\Lambda$
  increases.}
\end{table}

\begin{figure}
  \includegraphics[width=0.5\textwidth]{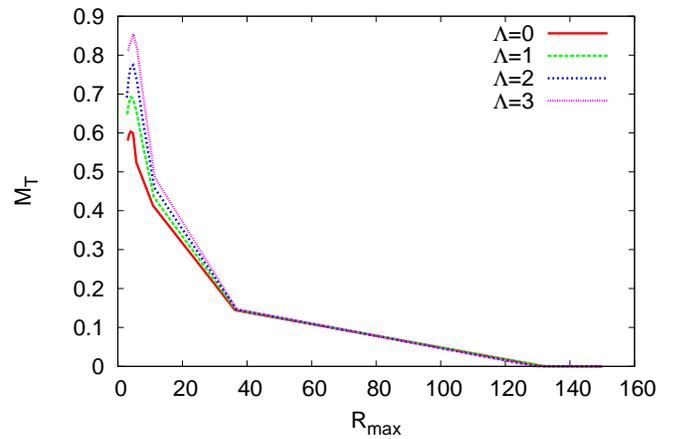}
  \caption{Typical graph of $M_{\Phi}$ in terms of the maximum
    radius $R_{max}$, the radius at which the radial metric function
    $g_{rr}$ reaches its maximum value, see
    Fig.~\ref{funciones}, for different values of $\Lambda$. The
    configurations become more massive and more compact for larger
    values of the self-interacting parameter.}
  \label{max}
\end{figure}

For completeness, we show in Fig.~\ref{max}, for different equilibrium
configurations, the critical mass $M_{\Phi c}$ as a function of the
radius $R_{max}$ at which the metric radial function reaches its
maximum value at $t=0$.  

\section{\label{sec:numer-evol-self}Numerical Evolution of
  self-interacting oscillatons}
  
We proceed now to the evolution of the EKG equations using the method
of lines. For the time integrations, we use a fourth order {\it
  Runge-Kutta} method (RK4), and second order centered differences to
discreticize spatial derivatives. A second-order {\it Runge-Kutta}
method (RK2) is used for the spatial integration of the metric
functions at each time level. Special care is needed for
Eq.~(\ref{set2}), otherwise its discretization would not be
second-order accurate because of the presence of the factor $1/x^{2}$
in the principal part. The transformation that guarantees the accuracy
of our simulations at the origin is
\begin{equation}
  \Pi_{,t} = 3 \frac{\partial}{\partial x^{3}} \left( \frac{x^{2}
      \alpha \Psi}{a} \right) - a \alpha ( \Phi + \Lambda \Phi^{3} )
  \, . \label{cambio}
\end{equation}

\subsection{\label{sec:boundary-conditions-1}Boundary Conditions}

To properly account for the origin at $x = 0$, we use the fictitious
point $x_{0}=-\Delta x/2$ as in \cite{Alcubierre:2003sx}, and take a
spatial grid of the form $x_{i} = (i-1/2)\Delta x$. From the evolution
equation of $\Phi$, Eq.~(\ref{set1}), we can see that it not necessary
to apply boundary condition for $\Phi$, because the evolution equation
can be integrated all the way from the boundary point $x_{0} =
-\Delta/2$ up to the outer boundary point. This is possible since the
evolution equation~(\ref{set1}) does not have spatial
derivatives. Likewise, we use the fictitious point to impose
appropriate parity conditions on the scalar field variables: $\Pi$ is
even, and $\Psi$ is odd.

At the outer numerical boundary, we assume that $\Pi$ behaves as an
outgoing wave pulse of the form
\begin{equation}
  \Pi=u(x-t)/x \, , \label{boundary1}
\end{equation}
where $u$ is an arbitrary function. In differential form,
Eq.~(\ref{boundary1}) becomes
\begin{equation}
  \partial_{x}\Pi +\partial_{t}\Pi +\Pi/x=0 \, . \label{diferencia}
\end{equation}
By using finite difference in Eq. (\ref{diferencia}), we can solve it
to find the unknown boundary value at the new time level. Because
$\Pi$ behaves as an outgoing wave at the boundary, so does
$\Phi$. Then, the outgoing wave boundary condition applied to $\Phi$
implies that at the outer boundary
\begin{equation}
  \Psi = -\Pi - \Phi/x \, . \label{condicionpsi}
\end{equation}
We used this expression to obtain boundary values for $\Psi$ after the
calculation of those of $\Phi$ and $\Pi$.

Boundary conditions for the metric functions are the following. Local
flatness at the origin implies that $a(x = 0) = 1$ and
$\partial_{x}a(x = 0)= 0$, and these two conditions together imply
that $a(x_{0}) = a(x_{1}) = 1+\mathcal{O}(\Delta x)^{3}$. We use these
two values at the first and second grid points to integrate the second
order Hamiltonian constrain outwards to obtain $a(x)$.

As for the lapse function, we impose $\alpha = 1/a$ as an outer
boundary condition; this is because in vacuum our slicing condition
implies that we are in Schwarzschild coordinates. Then, we are
assuming that our boundary conditions are sufficiently far away as to
be always in vacuum. Finally, the slicing condition is integrated
inwards to obtain the full profile of $\alpha(x)$.

\subsection{\label{sec:initial-conditions}Initial Conditions}
For most of our numerical experiments, we consider as initial
conditions the equilibrium configurations calculated in
Sec.~\ref{sec:equil-conf}; hence,
\begin{eqnarray}
  \Phi(t=0,x) &=& \sum_{j=1}^{J_{max}} \phi_{j}(x) \,
  , \label{condicion1} \\
  \Psi(t=0,x) &=& \Phi^{'}(0,x)= \sum_{j=1}^{J_{max}}
  \phi_{j}^\prime(x) \, , \label{condicion2} \\ 
  \Pi(t=0,x) &=& \dot{\Phi(0,x)}=0 \, . \label{condicion3}
\end{eqnarray}
Our interest resides mainly in the equilibrium properties of these
configurations, and it is first necessary to develop different
analysis techniques for their study. In this we will include, to begin
with, initial configurations that are a deformation of the original
equilibrium ones. There is a practical reason behind: equilibrium
configurations are easy to evolve, whereas general configurations may
be very difficult to follow as they develop in time.

\subsection{\label{sec:numerical-results-1}Numerical Results}
We start by calculating some properties of the equilibrium
configurations found now by the evolution code. For instance, the
critical mass $M_{\Phi c}$ of self-interacting oscillatons, for different
values of $\Lambda$, are shown in Table~\ref{tablados}.

\begin{table}[htp]
\begin{tabular}{ccc}
\hline
$\Lambda$ & $M_{\Phi c}$($m^{2}_{Pl}/m_{\Phi}$) & $\phi_{1 c}(0)$ \\  
\hline
0.0 & 0.599 & 0.47 \\ 
1.0 & 0.686 & 0.46 \\ 
2.0 & 0.767& 0.42 \\ 
3.0 & 0.843 & 0.39\\
\end{tabular}
\caption{Resulting values of the critical mass $M_{\Phi c}$ for the 
  numerical evolution of equilibrium oscillatons using different
  $\Lambda$'s, see also Table~\ref{masafrec} and text below for
  details.} 
\label{tablados}
\end{table} 

The total mass was obtained through the integration of the energy
density calculated from the time-time component of the scalar field
stress-energy tensor, $\rho_{\Phi} = -{T^{0}}_{0}$. Because of the
spherical symmetry, this method gives identical results to the
Schwarzschild mass in Eq.~(\ref{eq:mass}). We notice a little
discrepancy with respect to the values reported in
Sec.~\ref{sec:equil-conf}, this is because of the evolution code is
more accurate and does not depend upon the truncation we had to impose
upon the Fourier expansions in~(\ref{eq:fourier}).

\subsubsection{\label{sec:code-tests}Code tests}
\begin{figure}[htpb!]
  \includegraphics[width=0.49\textwidth]{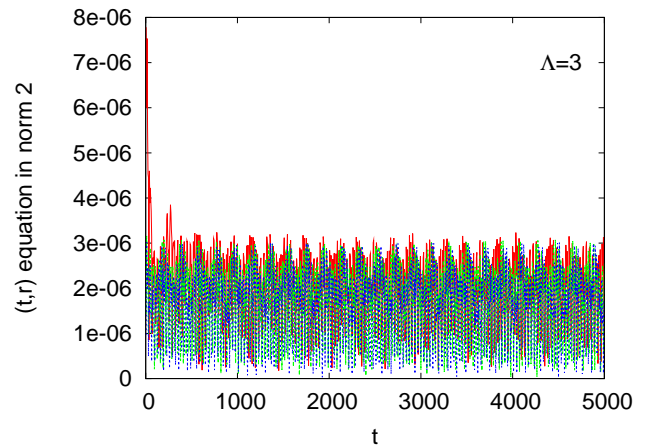}
  \caption{$L2$-norm of the momentum constraint~(\ref{einstein1}) for
    three different resolutions $dx=0$.$01$ (red), $dx=0$.$02$
    (green), $0$.$04$ (blue) and $\Lambda= 3$. It can be seen that the
    accuracy of the code is kept under control along the runs.}
    \label{conver}
\end{figure}

To revise the accuracy of our numerical calculations, we monitored the
momentum constraint~(\ref{einstein1}), 
\begin{equation}
\beta:= a_{,t}-\frac{1}{2}x\alpha\Psi\Pi=0 \, , \label{convergencia}
\end{equation}
and calculated the $L_2$-norm of the value of $\beta$ across the
grid as a function of time using three different spatial grid sizes,
see Fig.~\ref{conver}. It can be seen that the accuracy of the
numerical code is not altered by the values of the quartic parameter
$\Lambda$.

\begin{figure}[htbp!]
  \includegraphics[width=0.49\textwidth]{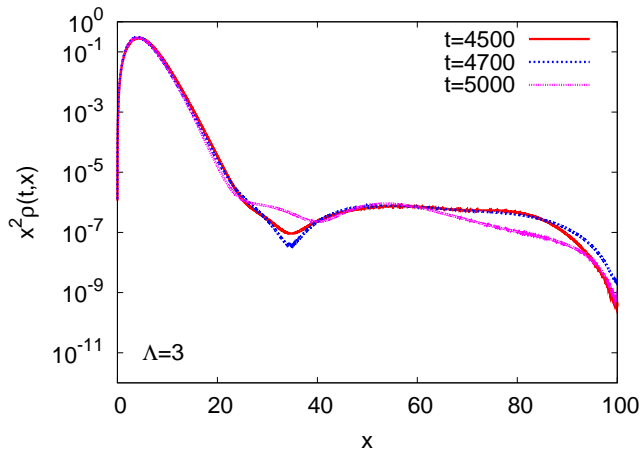}
  \caption{The function $x^{2}\rho$ in the case $\Lambda= 3$. The
    small discrepancies seen at the outer parts of the numerical
    domain, of the order of $x^{2}\rho < 10^{-5}$, appear because of
    the boundary condition used, which is, in the strict sense, only
    appropriate for the massless case. However, the numerical error is
    kept under control all along the runs.}
    \label{densidad}
\end{figure}

On the other hand, Fig.~\ref{densidad} shows the mass-differential
function $x^{2}\rho$ for a run with a numerical boundary located at
$x_{max}=100$, and for $\Lambda =3$. We can see that the plots are
practically the same up to $x=20$, but differ one from each other at
the outer parts of the numerical domain. These discrepancies, of the
order of $x^{2}\rho < 10^{-5}$, arise because part of scalar field has
been reflected from the outer boundary. The error is not significant
and we were able to keep it under control during the numerical
evolution.

\begin{figure}[htbp!]
  \includegraphics[width=0.49\textwidth]{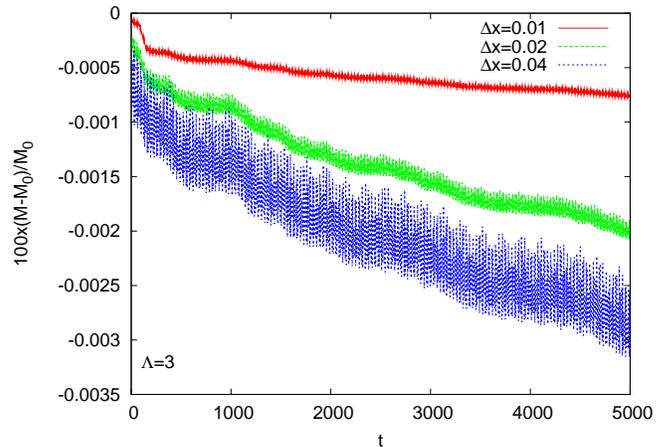}
  \caption{Typical case of mass dissipation in the numerical evolution
    of equilibrium configurations. It is caused by the numerical error
    in the evolution, but we confirmed in our numerical experiments
    that can be reduced by considering a finer spatial grid.}
    \label{difermass}
\end{figure}

Fig.~\ref{difermass} shows the evolution of the total integrated mass
for each time level. We can observe a small adjustment of the original
 mass, because there is a small ejection of scalar field at the
 beginning of the run. We can notice a steady decay of the mass, which
 is evidence of an intrinsic dissipation of our numeric code; it can
 be reduced by taking a finer spatial grid,
 and then it cannot be considered an intrinsic decay of the
 oscillatons.

\subsubsection{\label{sec:s-branch-quasi}S-branch and Quasi-normal Modes}

We present here numerical evidence that the characterization of
stable, S-branch, and unstable, U-branch oscillatons is preserved in
the case of the inclusion of a quartic interaction, much in the same
manner as in the case of non self-interacting oscillatons and boson
stars.

\begin{figure}[htbp!]
  \includegraphics[width=0.49\textwidth]{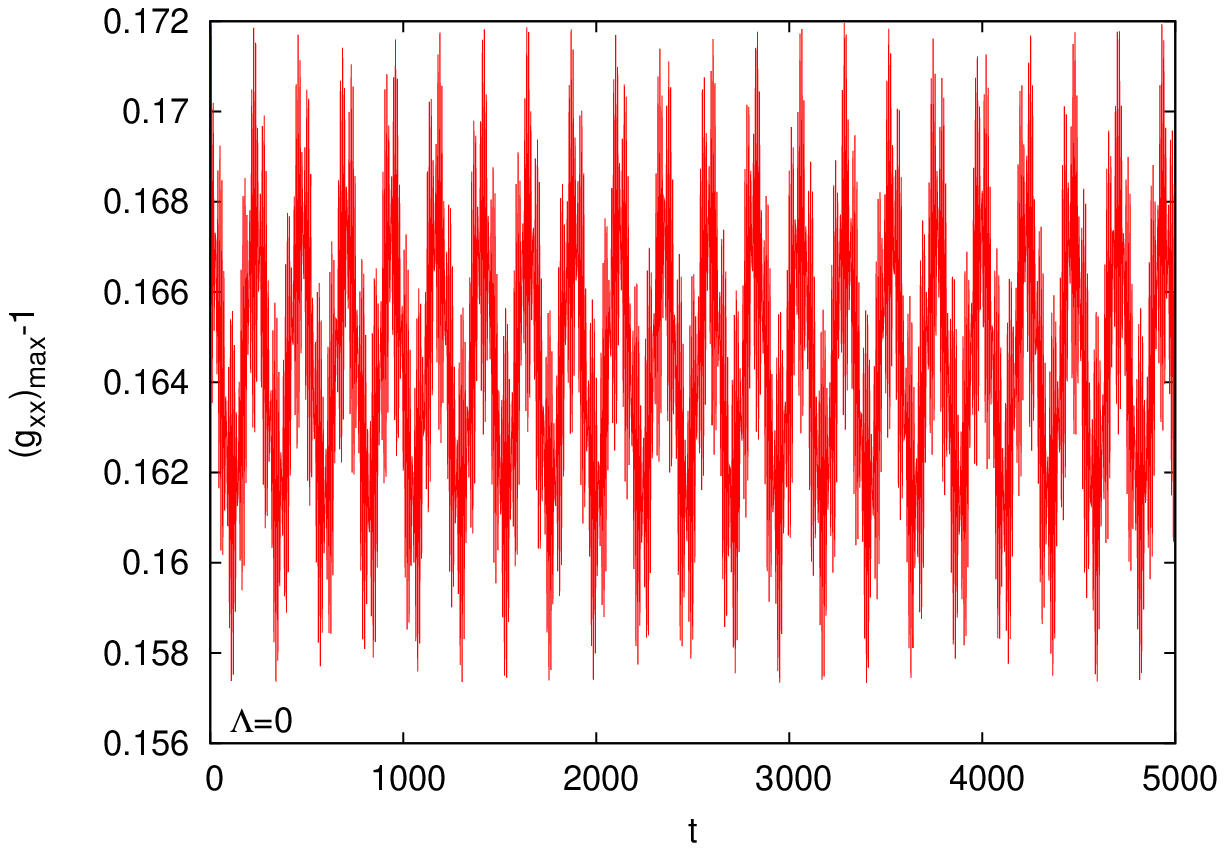}
  \includegraphics[width=0.49\textwidth]{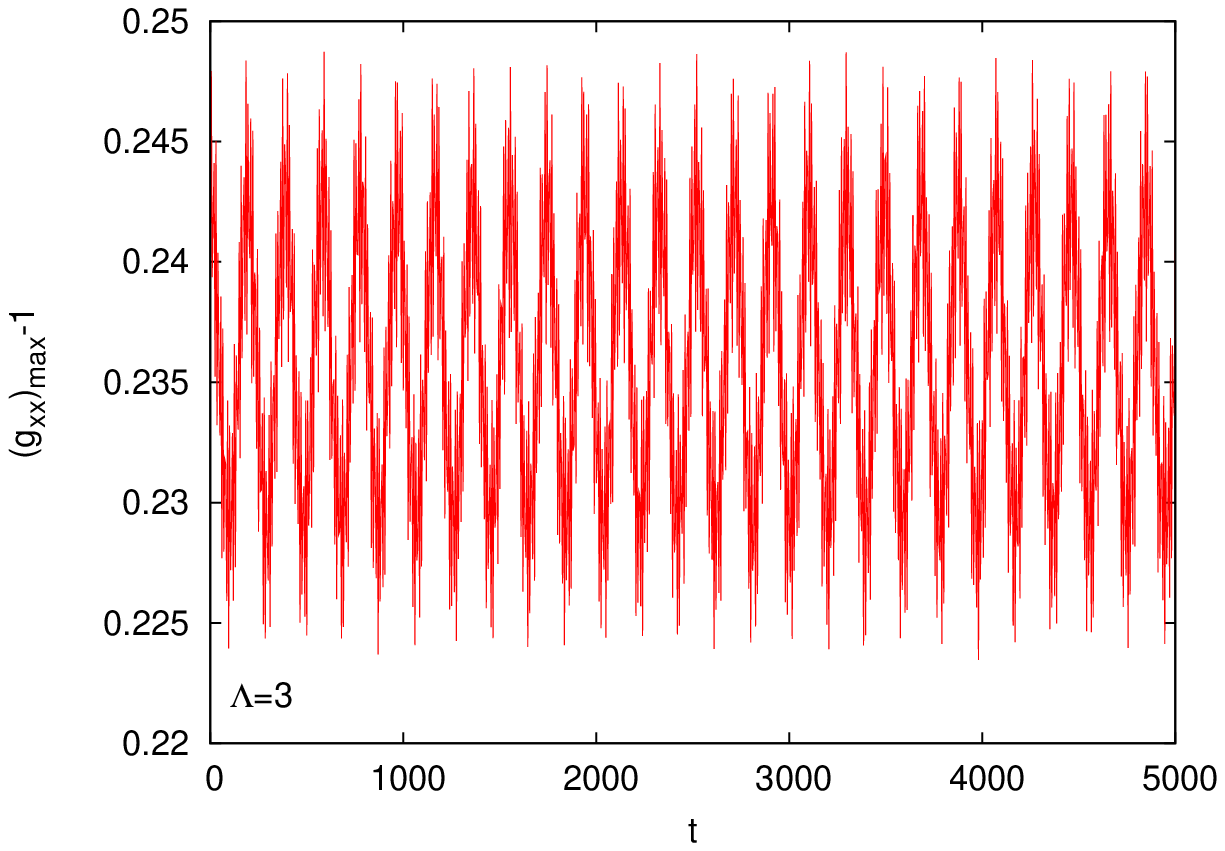}
  \caption{Maximum value of the radial metric function
    $g_{rr}(t,x)=a^{2}(t,x)$, corresponding for the initial
    configuration $\phi_{1}(0)=0$.$28$ with $\Delta x=0.01$ and 
    $\Delta t=0.005$, at $x_{max}=100$ and the evolution is shown to
    $t=5000 m^{-1}$ for $\Lambda=0$ (Top) and $3$ (Bottom).
  }
    \label{rmax}
\end{figure}

To begin with, we show in Fig.~\ref{rmax} the evolution of the maximum
value of the radial metric function $g_{xx}=a^{2}(t,x)$, corresponding
to an equilibrium configuration with $\phi_{1}(0)=0.28$. The outer
boundary of the numerical domain was set at $x_{max}=100$, and the run
was followed up to a time $t=5000$. It can be noticed that the
configuration maintains the same oscillatory pattern at all times and
for all values of $\Lambda$. This is an evidence of the stability of
the oscillations in response to small radial perturbations. In this
case, the perturbations come from the truncation of the Fourier
series~(\ref{eq:fourier}) and the discretization error of the
numerical solutions.

\begin{figure}[htbp!]
  \includegraphics[width=0.49\textwidth]{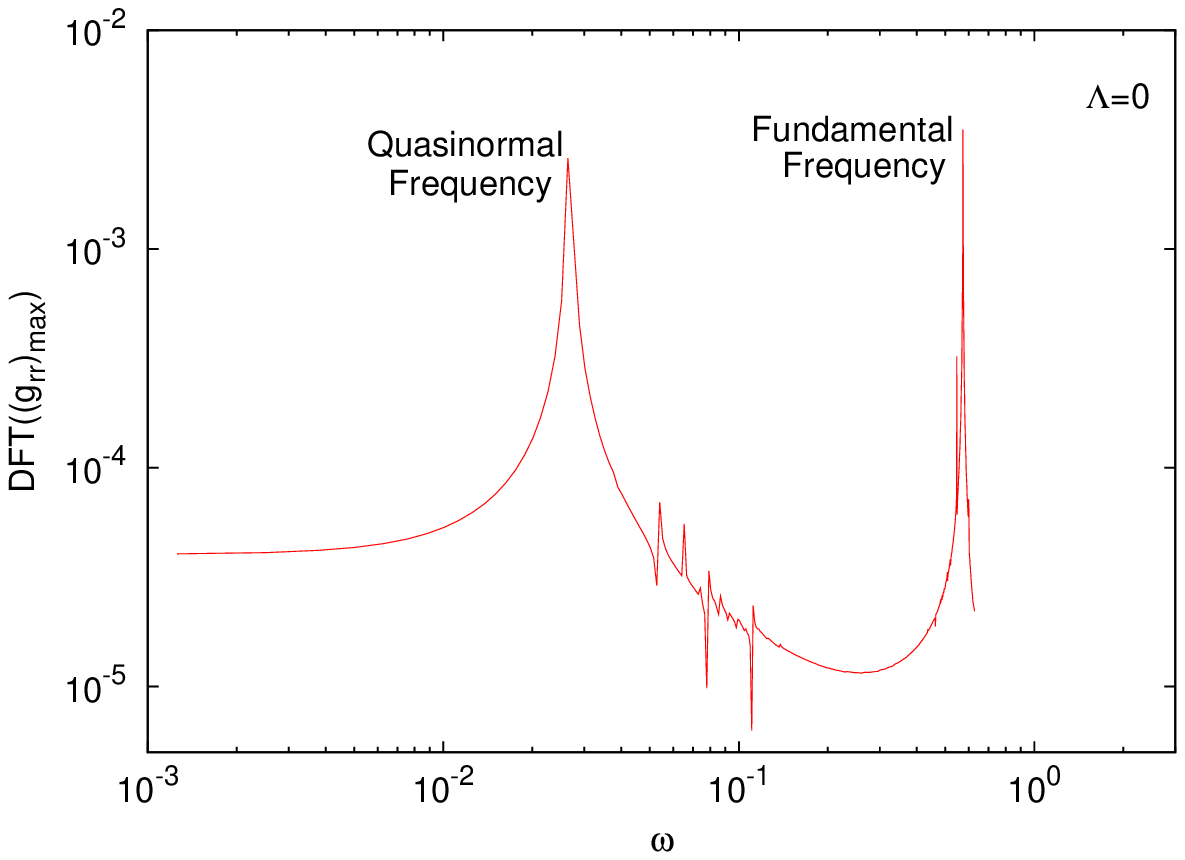}
  \includegraphics[width=0.49\textwidth]{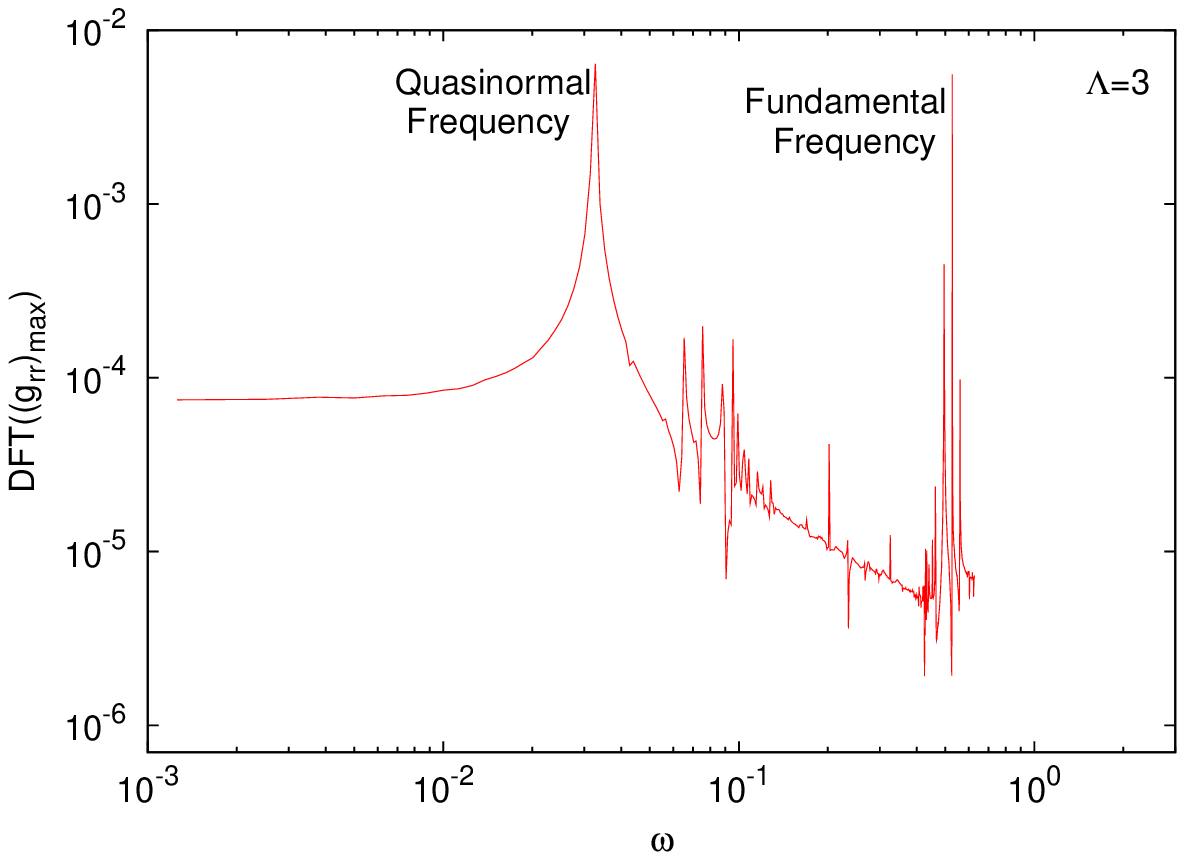}
  \caption{\label{Fourier}We show the Fourier transform of the evolution
    of the maximum value of the metric coefficient $g_{rr}$ for the
    configuration with $\phi_{1}=0.28$. The corresponding values of
    the quasi-normal frequencies for the different values of $\Lambda$
    are: $(f/m)=4.2 \times 10^{-3}$ ($\Lambda=0$), $(f/m)=4.8 \times
    10^{-3}$ ($\Lambda=1$),  $(f/m)=5.0 \times 10^{-3}$ ($\Lambda=2$),
    $(f/m)=5.05 \times 10^{-3}$ ($\Lambda=3$), while their fundamental
    frequencies are: $(\omega/m)=0.91$ ($\Lambda=0$),
    $(\omega/m)=0.90$ ($\Lambda=1$), $(\omega/m)=0.89$ ($\Lambda=2$)
    and $(\omega/m)=0.88$ ($\Lambda=3$). Quasi-normal frequencies
    increase with $\Lambda$, a property that can also be noticed in
    Fig.~\ref{rmax}.
}    
\end{figure}

Fig.~\ref{Fourier} presents the Fourier transform of the
oscillations shown in Fig.~\ref{rmax}; we observe that the maximum of
the $g_{rr}$ oscillates periodically with two distinctive time
scales. The short-period oscillation (large frequency) corresponds to
the fundamental frequency used in the Fourier
expansions~(\ref{eq:fourier}). The large-period oscillation (small
frequency) is an overall vibration of the configuration that we
identify as the characteristic \emph{quasi-normal} modes of the
oscillatons.

\begin{figure}[htbp!]
  \includegraphics[width=0.49\textwidth]{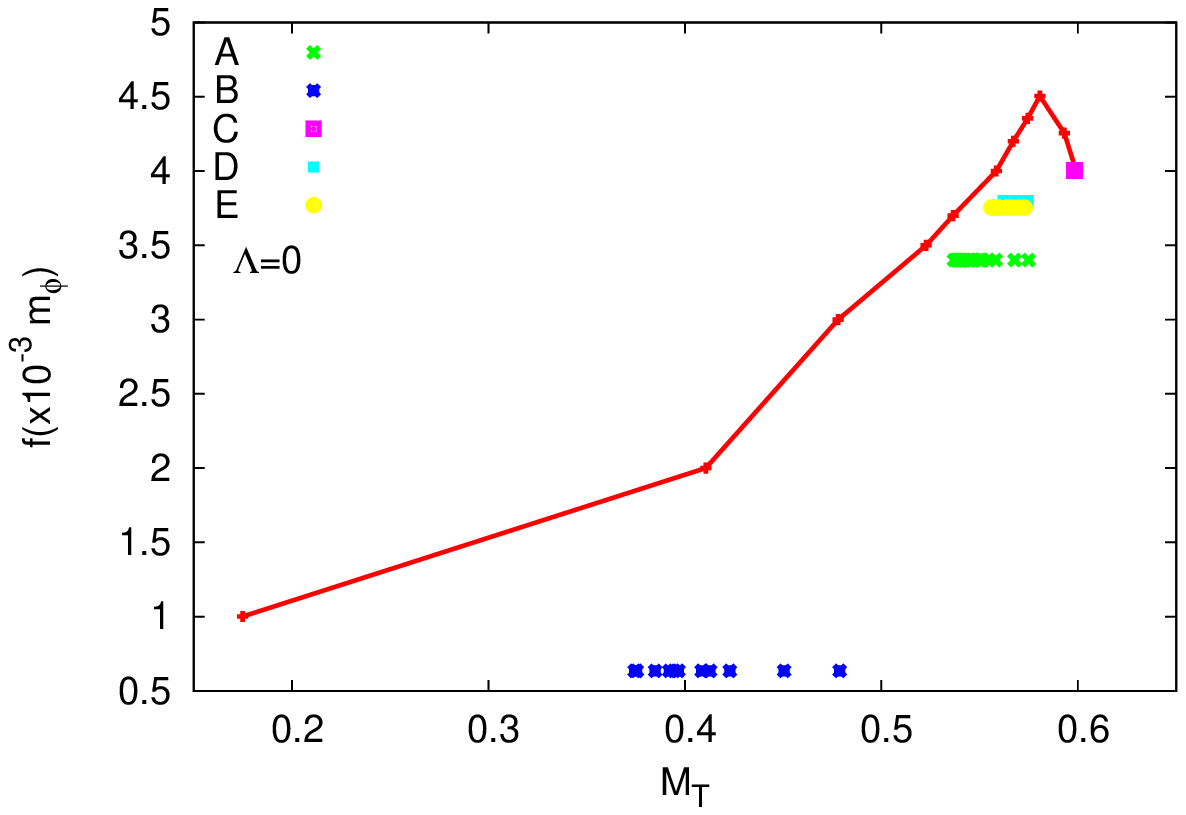}
  \includegraphics[width=0.49\textwidth]{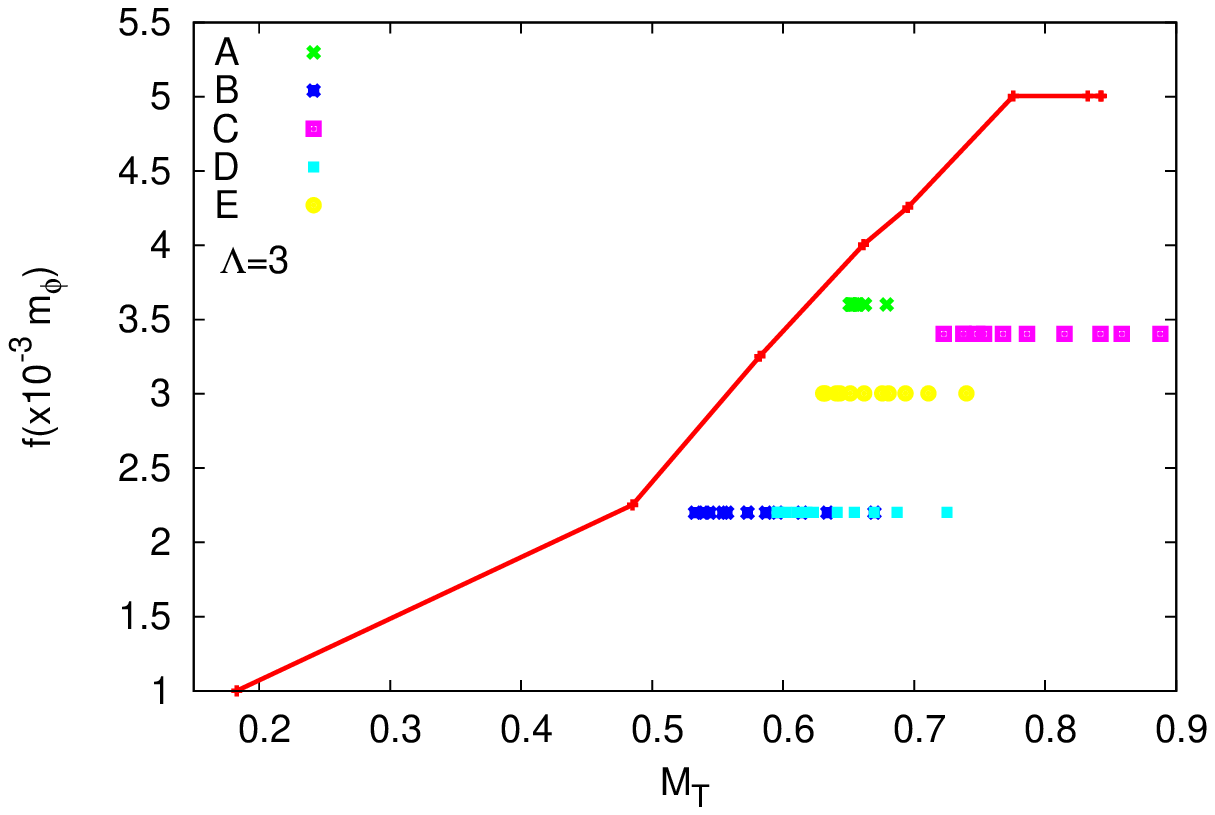}
  \caption{The quasi-normal frequencies obtained from the evolution
    of slightly perturbed S-branch oscillatons as a function of the
    total mass $M_T$. Also shown is the migration of perturbed
    S-branch and U-branch oscillatons, labelled $A-E$, see Figs.}
    \label{quasi}
\end{figure}

Following the analysis done in \cite{Alcubierre:2003sx}, we
calculate the power spectrum of the evolution for the entire S-branch;
we show in Fig.~\ref{quasi} the quasi-normal frequency $f$ as a
function of the total mass $M_{T}$. This kind of plots has proved
useful in the analysis of the evolution of general scalar field
configurations,
see\cite{Seidel:1990jh,Balakrishna:1997ej,Alcubierre:2003sx}.

\begin{figure}[htbp!]
  \includegraphics[width=0.49\textwidth]{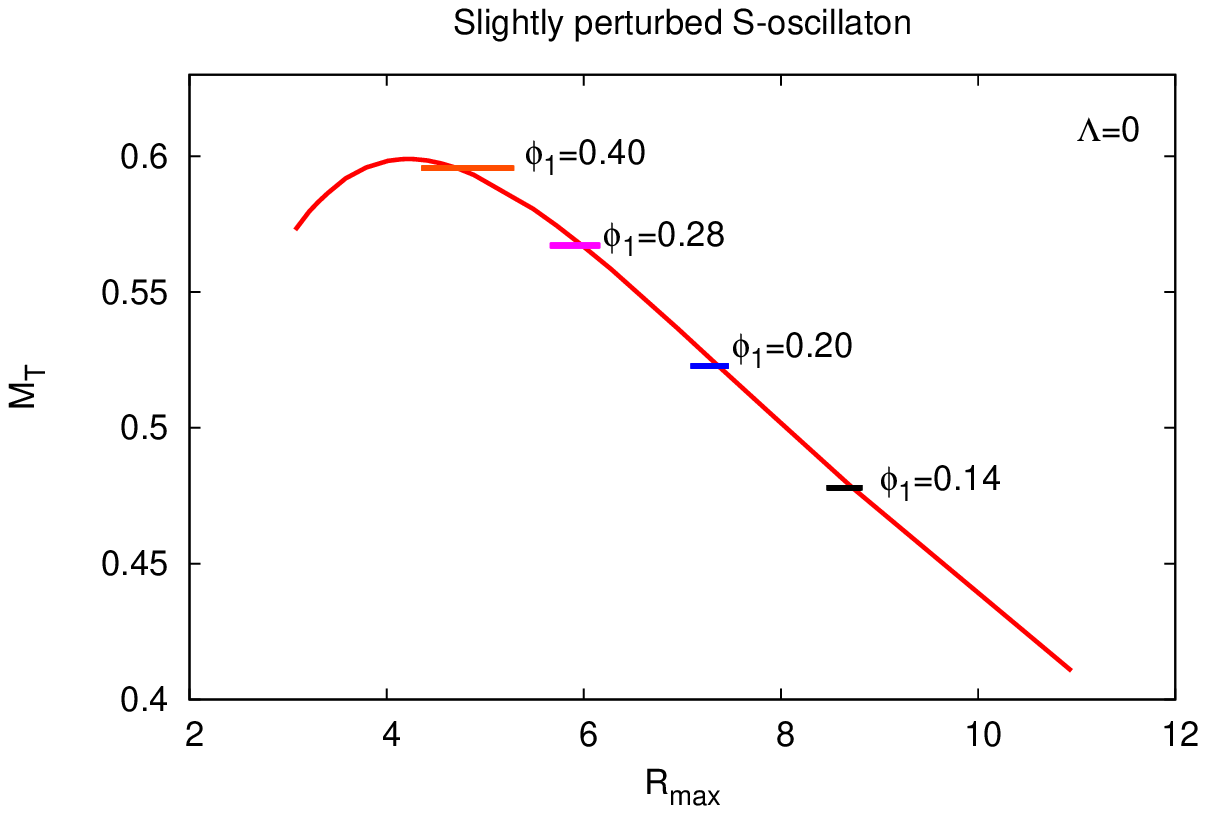}
  \includegraphics[width=0.49\textwidth]{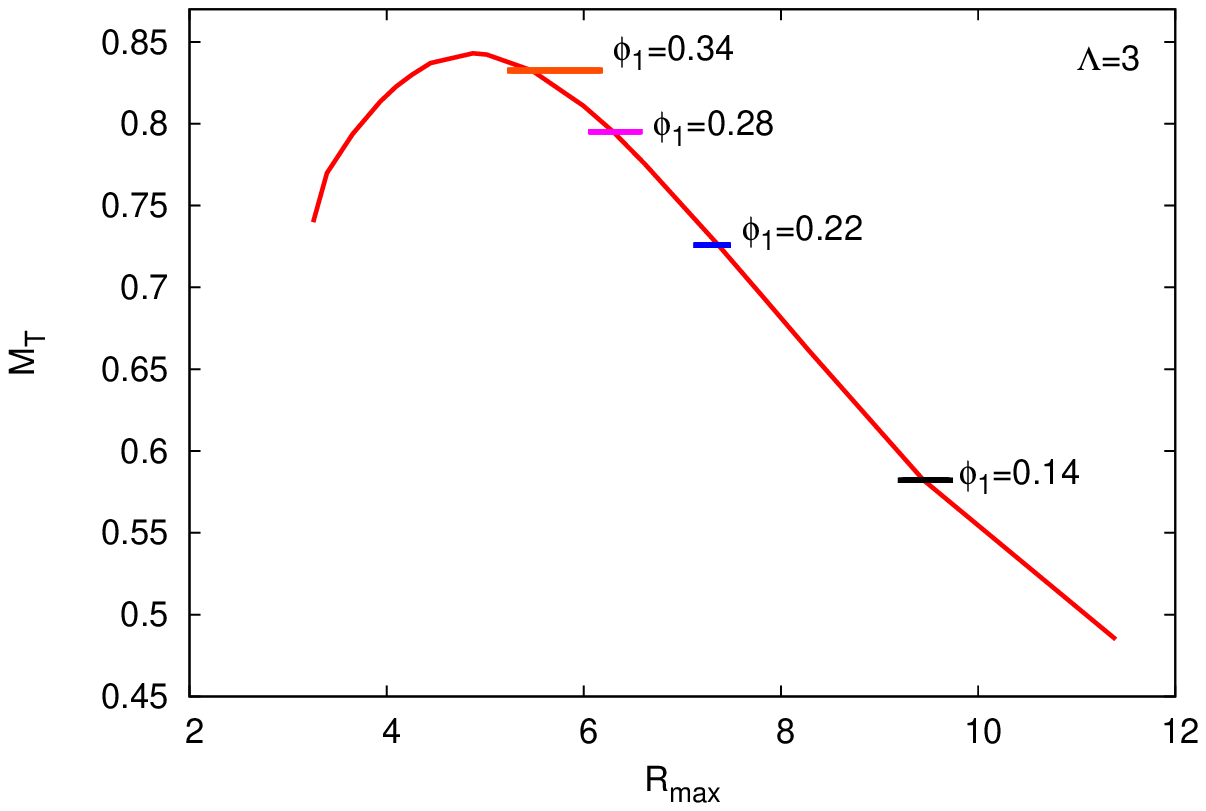}
  \caption{\label{masse}We show the evolution of the total mass $M_{T}$ and
    $R_{max}$ for different slightly-perturbed S-oscillatons up to a
    time $t=5000 \, m^{-1}$. S-oscillatons are not migrating nor
    decaying, but only oscillating around their equilibrium positions
    in this plot.
}
\end{figure}

On the other hand, Fig.~\ref{masse} presents the evolution of the
total mass $M_{T}$ and $R_{max}$ ($R_{max}$ is the value where the
radial metric function $g_{rr}$ reaches its maximum value) as compared
with the values corresponding to equilibrium configurations. All the
cases correspond to the S-branch.

It is observed that slightly perturbed S-oscillatons are not migrating
to another S-oscillaton profile, but rather they oscillate with a
small amplitude around the original equilibrium configuration. It is
then concluded that S-oscillatons are stable under small radial
perturbations, and that the frequencies show in Fig.~\ref{Fourier} are
their intrinsic quasi-normal modes.

\subsubsection{\label{sec:pert-s-oscill}Perturbed S-oscillatons}
\begin{figure}[!htbp]
  \includegraphics[width=0.49\textwidth]{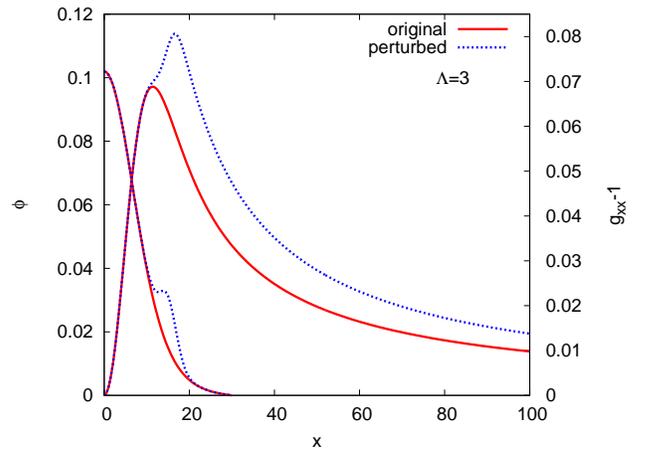}
  \caption{\label{gauss}The scalar field $\Phi$ and the radial metric
    function $g_{rr}$ profiles of a strongly perturbed S-oscillatons with
    $\phi_{1}(0)=0$.$1$ for the case with a mass increase of $40$
    percent.
}
\end{figure}

 We now turn our attention to the evolution of some strongly-perturbed
 S-oscillatons. We use a Gaussian profile as a perturbation applied to
 the original equilibrium configurations; see Fig.~\ref{gauss} for an
 example of this perturbation in the case of an S-oscillaton with
 $\phi_{1}(0)=0.1$, so that its mass is increased by $40$\%. The purpose
 is to analyze whether S-oscillatons are stable under strong
 perturbations, and the conditions to be met for the collapse into a
 black hole. 
 
First, we increased the original mass of the equilibrium configuration
(see Fig.~\ref{masas}) with $\phi_{1}(0)=0.1$ by $40$\% and $60$\% for
all $\Lambda$ values. The initial masses for $\Lambda=1$ are
$M_{i}=0.5748$ ($40\%$) and $M_{i}=0$.$6816$ ($60\%$), for $\Lambda=1$
are $M_{i}=0.6075$ ($40\%$) and $M_{i}=0.7203$ ($60\%$), for
$\Lambda=2$ are $M_{i}=0.6423$ ($40\%$) and $M_{i}=0.7615$ ($60\%$),
and for $\Lambda=3$ are $M_{i}=0.6741$ ($40\%$) and $M_{i}=0.8051$
($60\%$). 

\begin{figure}[htbp!]
  \includegraphics[width=0.49\textwidth]{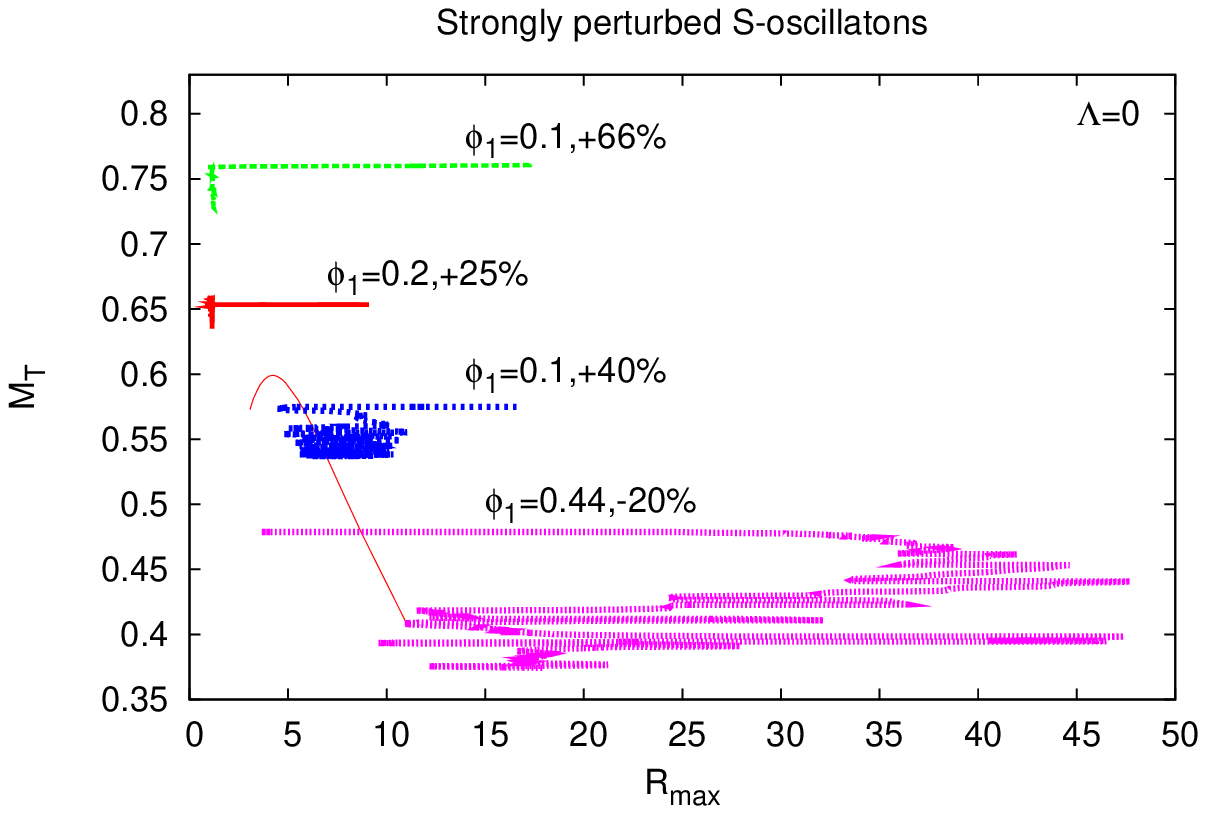}
  \includegraphics[width=0.49\textwidth]{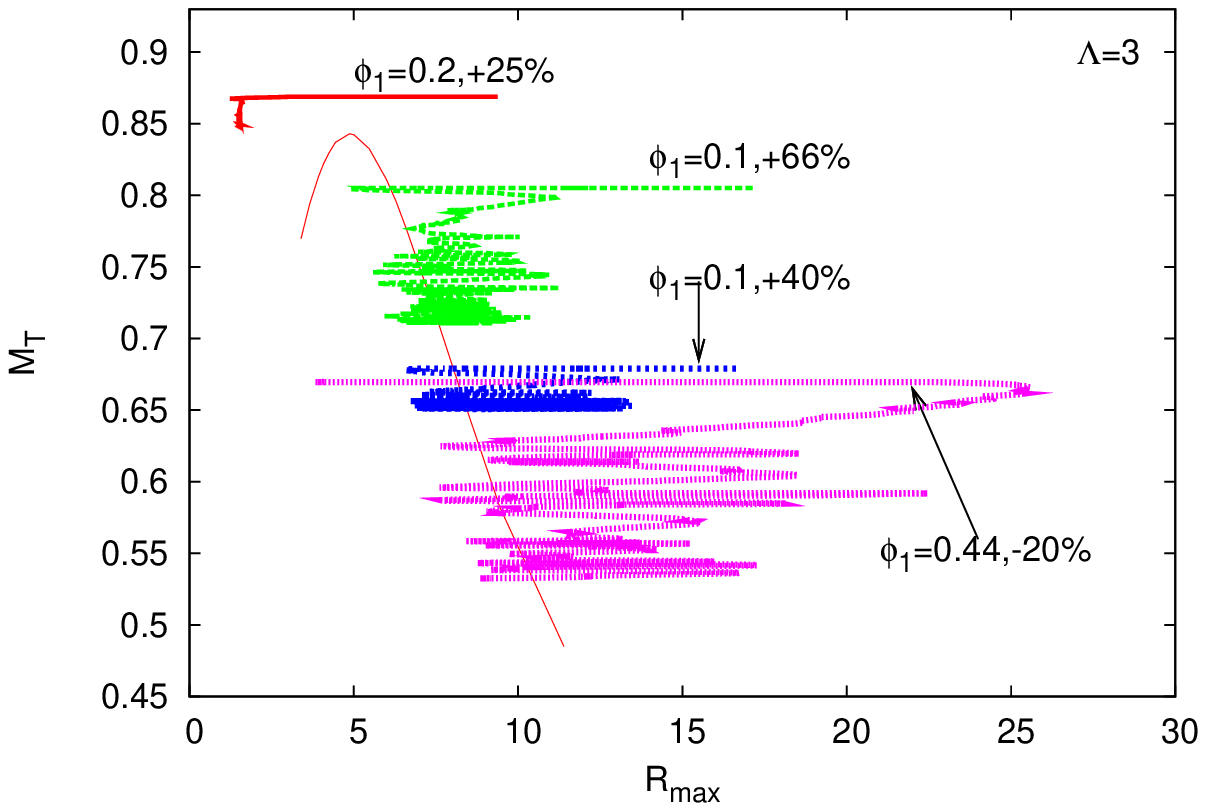}
  \caption{Evolution of different strongly perturbed S-oscillatons. If
  the initial mass is less than the critical value $M_{\Phi c}$, the
  configurations are able to evolve towards another
  S-configuration. Otherwise, the final fate may be the collapse into
  a black hole.}
    \label{sp}
\end{figure}

From Fig.~\ref{sp} we can see that for $\Lambda=0$ the perturbed
configuration with its mass increased by $60$\% collapses into a black
hole, while for the rest of the $\Lambda$ values this configuration
migrates to another oscillaton located on the S-branch. The reason for
this is that the $\Lambda=0$-configuration has an initial mass that is
larger than the critical mass of equilibrium configurations (see
Table~\ref{masafrec}), whereas for the other cases the initial mass is
smaller than the critical one. The perturbed configurations with their
mass increased by $40\%$ are able to migrate to another S-oscillaton. 

For equilibrium configurations with $\phi_{1}(0)=0.2$, we increased
its original mass in $25\%$ obtaining the following initial masses:
$M_{i}=0$.$6534$ ($\Lambda=0$), $M_{i}=0$.$7198$ ($\Lambda=1$),
$M_{i}=0$.$7919$ ($\Lambda=2$), and $M_{i}=0$.$8687$
($\Lambda=3$). This configuration collapses into a black hole for all
values of $\Lambda$; in all cases, the initial mass of the perturbed
configuration is larger than critical one corresponding to each case. 

Finally, for equilibrium configurations with $\phi_{1}(0)=0.44$, we
decreased its original mass by $20$\%: $M_{i}=0$.$4787$
($\Lambda=0$),$M_{i}=0$.$5484$ ($\Lambda=1$), $M_{i}=0$.$6133$
($\Lambda=2$), and $M_{i}=0$.$6696$ ($\Lambda=3$). As we can see in
Fig.~\ref{sp}, these configurations lose mass until they reach the
position of another S-oscillaton.

\begin{figure}[htbp!]
  \includegraphics[width=0.49\textwidth]{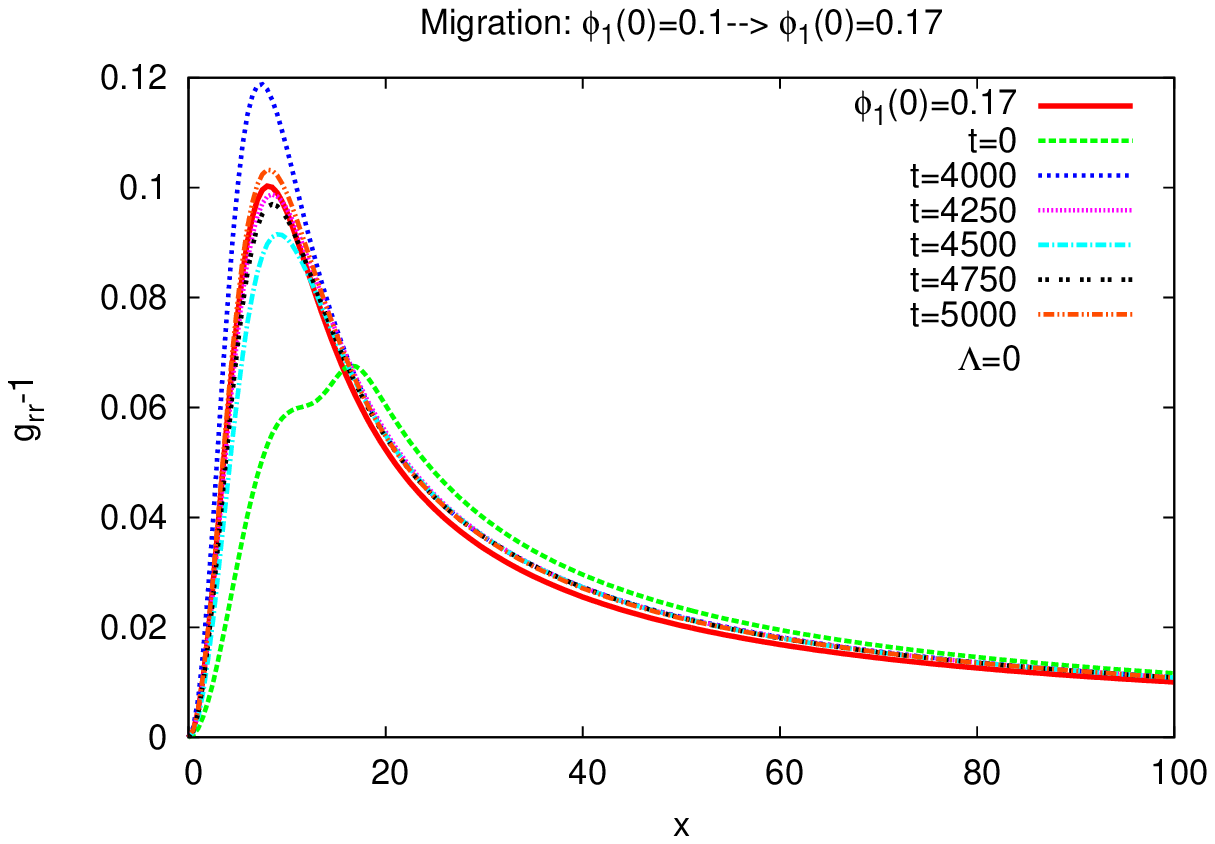}
  \includegraphics[width=0.49\textwidth]{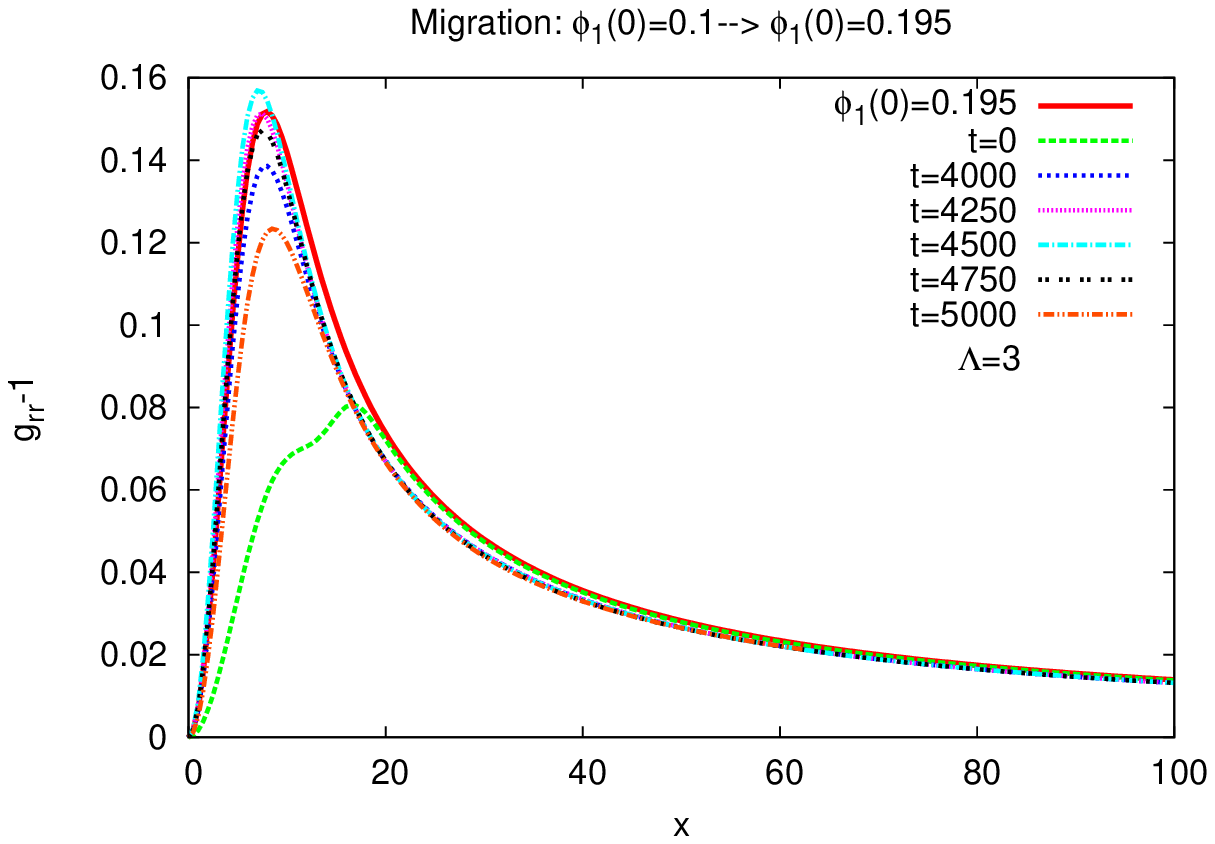}
  \caption{Evolved profiles of the metric coefficient $g_{rr}$ for a
    strongly perturbed $\phi_{1}(0)=0.1$-oscillaton for different
    values of $\Lambda$. Each plot indicates the final S-oscillaton
    each initial configuration is migrating to.}
 \label{migracion}
\end{figure}

We have noticed that oscillatons maintain a fixed vibration frequency
during its evolution. This is shown in Fig.~\ref{sp}, where we present
the migration path of oscillatons with $\phi_{1}(0)=0.1$ (with its
mass increased by $40$\%), and with $\phi_{1}(0)=0.44$ (with its mass
decreased by $20$\%), which are labeled $A$ and $B$
in Fig.~\ref{quasi}. We can appreciate that the perturbed oscillaton with
$\phi_{1}(0)=0.1$ ($+40\%$) is migrating to an equilibrium
configuration with $\phi_{1}(0)=0.17$. This can also can be seen in
Fig.~\ref{migracion}, which shows the profile of the metric
coefficient $g_{rr}$ rapidly approaching and oscillating around the
final configuration.

Then, as reported in Ref.~\cite{Alcubierre:2003sx}, we have found that
strongly perturbed S-oscillaton are able to migrate to another
S-oscillaton when their original mass is smaller than critical
mass. But, if the original mass increases enough to be larger than the
critical one, this perturbed configuration will collapse into a black
hole, except in the case of diluted oscillatons, with low values of
$\phi_0$, in which the collapse to a black hole can be prevent by the
gravitational cooling mechanism.   

\subsubsection{\label{sec:u-branch}U-branch}

We identify as U-oscillatons the equilibrium configurations that are
located on the right-hand side of the critical configuration in a plot
of $M_{T}$ versus $\phi_{1}(0)$, see Fig.~\ref{masas}, or located on
the left-hand side in a plot of $M_{T}$ versus $R_{max}$, see
Fig.~\ref{max}. To evolve these equilibrium configurations, we also
use the slightly-perturbed, by numerical inaccuracies, configurations,
which in general decay and migrate to a configuration located on the
S-branch.  

\begin{figure}[htpb!]
  \includegraphics[width=0.49\textwidth]{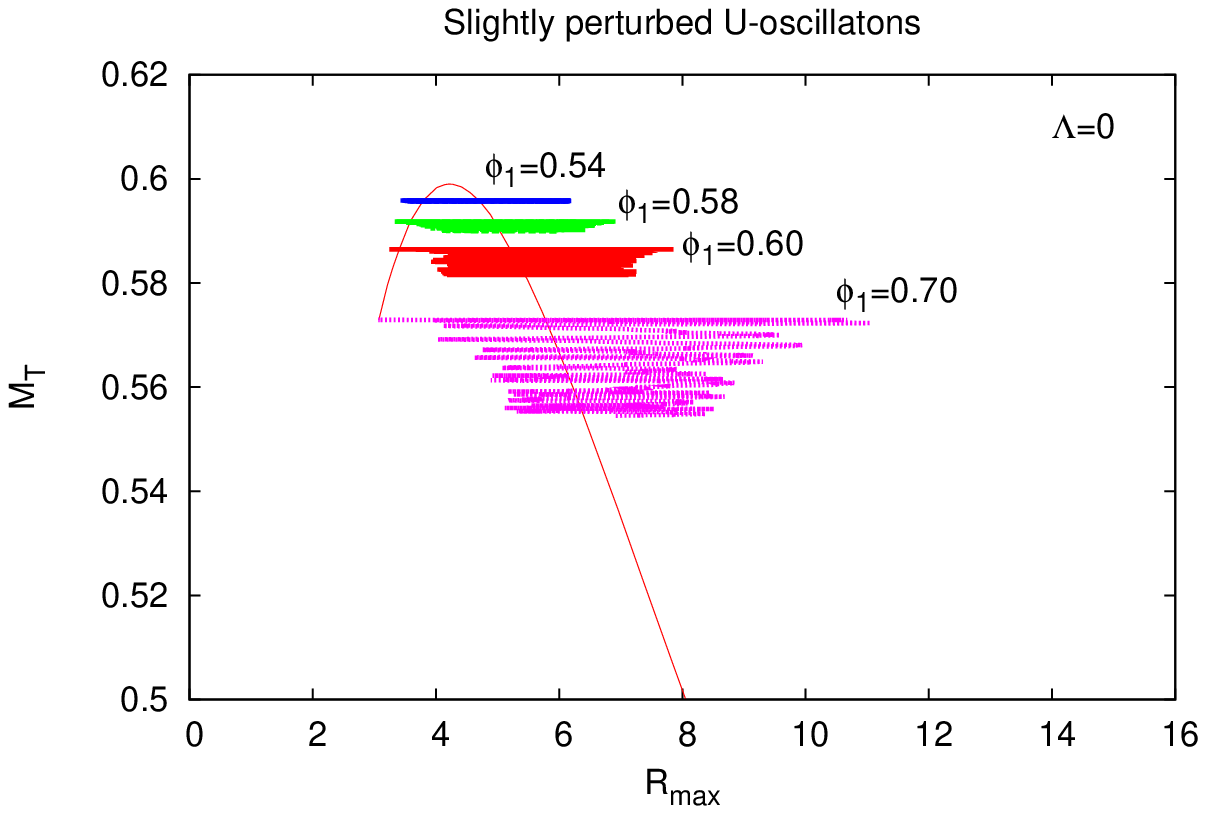}
  \includegraphics[width=0.49\textwidth]{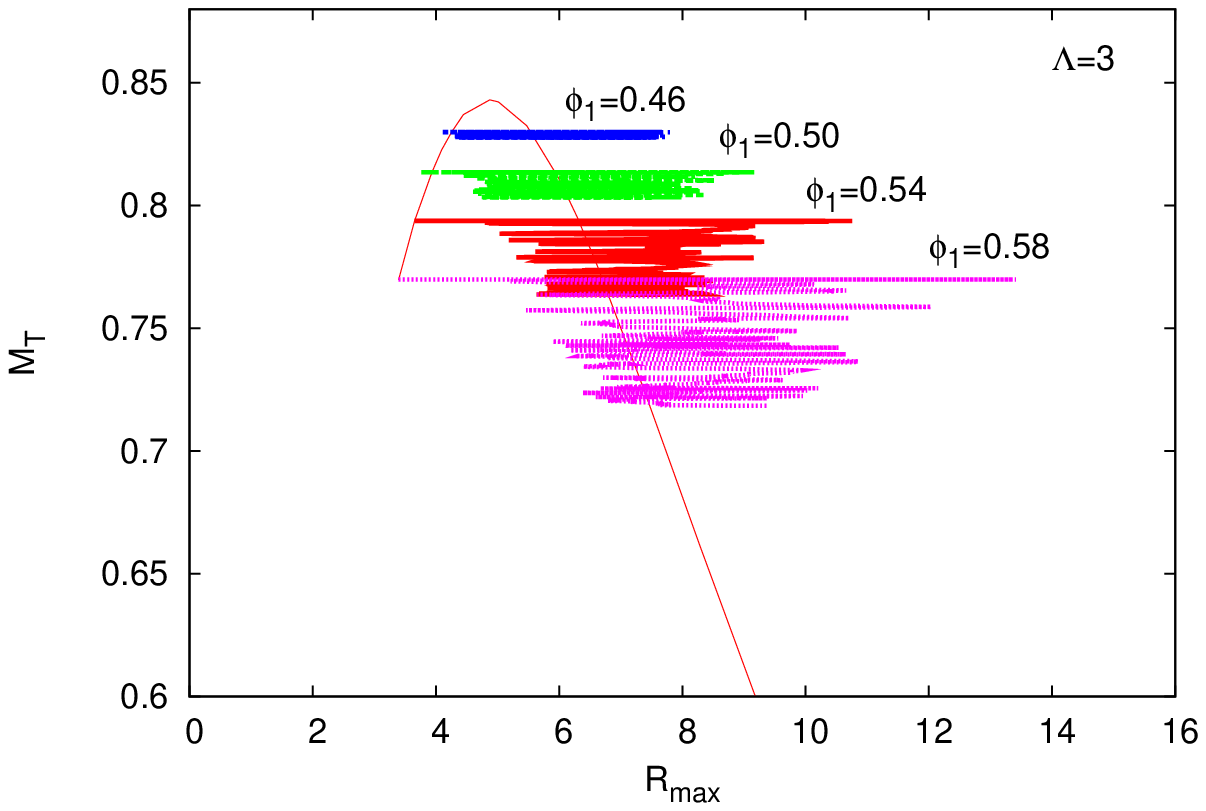}
  \caption{Slightly-perturbed U-branch configurations. In general, all
    of them migrate (left to right on the plots) to an equilibrium
    configuration located on the S-branch.}  \label{massu}
\end{figure}

\begin{figure}[htpb!]
  \includegraphics[width=0.49\textwidth]{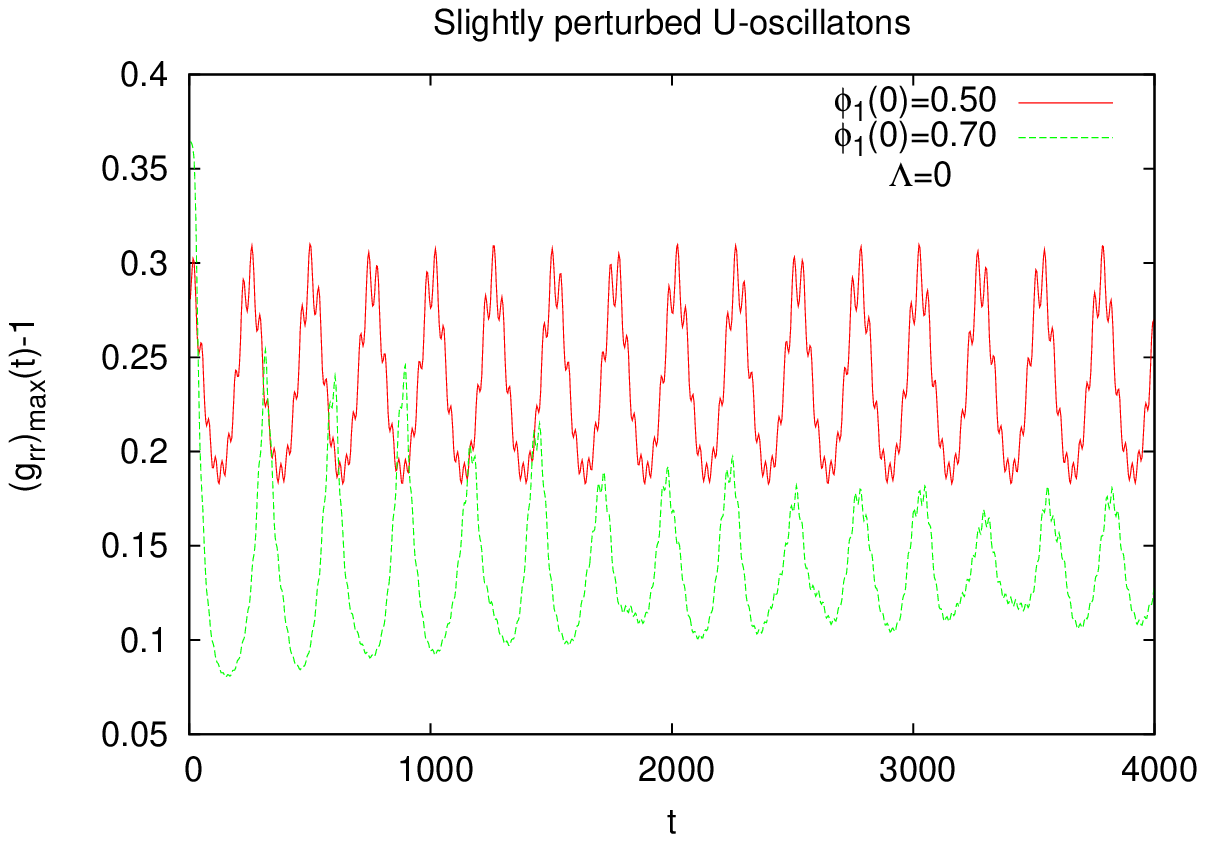}
  \includegraphics[width=0.49\textwidth]{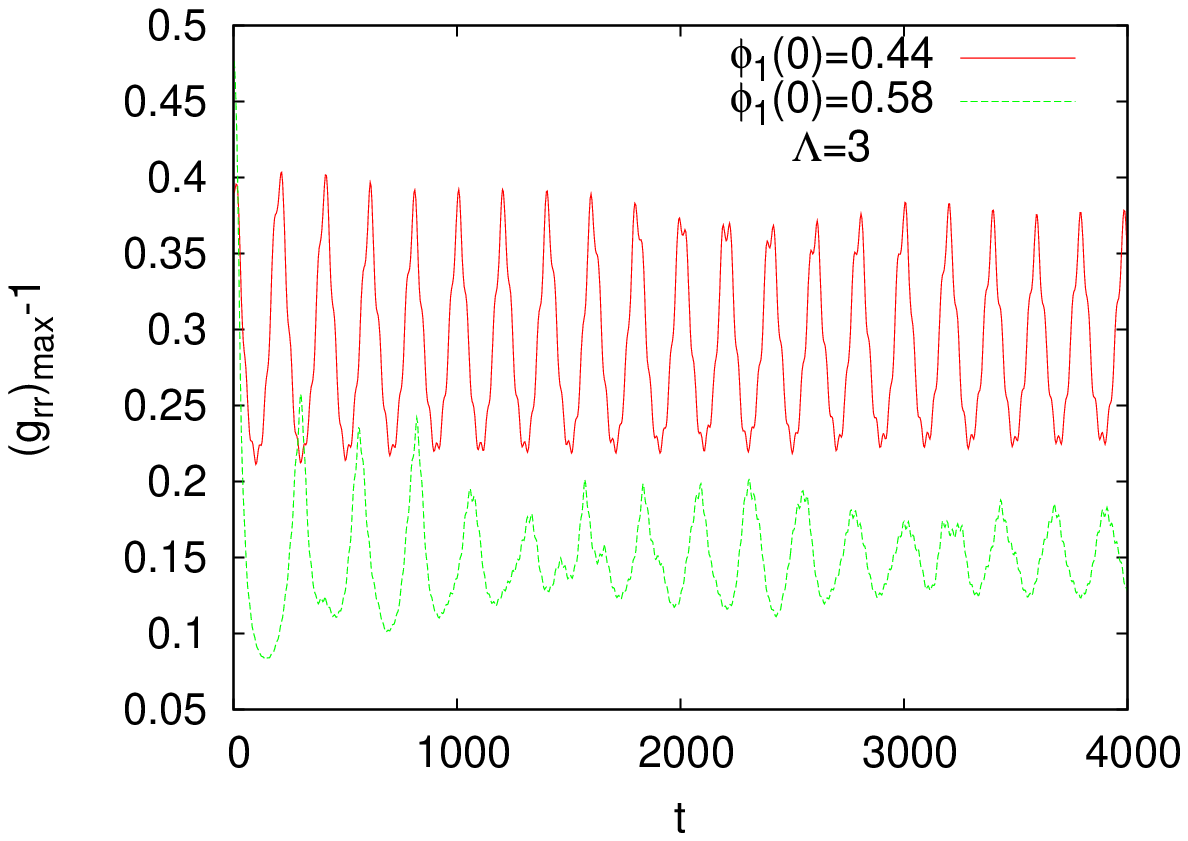}
  \caption{ Evolve profiles of the maximum value of the radial metric
    function $g_{rr}$ for the slightly perturbed configurations:
    $\phi_{1}(0)=0$.$5$ and $\phi_{1}(0)=0$.$7$} 
   \label{grrp}
\end{figure}

Fig.~\ref{massu} shows some slightly perturbed U-oscillatons, and we
can see that the equilibrium configuration is more unstable (the
quicker it starts to migrate) for larger values of $\phi_{1}(0)$. This
also can be seen in Fig.~\ref{grrp}, where we show the evolution of
the maximum value of  the radial metric coefficient $g_{rr}$ for the
slightly perturbed configurations with central field values
$\phi_{1}(0)=0.5$ and $\phi_{1}(0)=0.7$.

Thus, we confirm that the U-oscillatons are intrinsically unstable
under small perturbation, they decay and migrate to the S-branch. In
Fig.~\ref{quasi}, we show the migration of the slightly perturbed
U-oscillaton with $\phi_{1}(0)=0.5$ and $\phi_{1}(0)=0.7$, labeled
$C$ and $E$ for $\Lambda=0$, $\phi_{1}(0)=0.7$ labeled $E$ for
$\Lambda=1$ and $\Lambda=2$, and $\phi_{1}(0)=0.58$ labeled $E$ for
$\Lambda=3$.

\subsubsection{\label{sec:perturbed-u-branch}Perturbed U-branch}

We study now the behavior of the U-branch equilibrium configurations
under strong perturbations. An example of the typical behaviors is
provided by the configuration with $\phi_{1}=(0)=0.6$ for all values
of $\lambda$. First, we increase its original mass by $2$\%; the
resulting initial masses are:  $M_{i}=0.598$ ($\Lambda=0$),  $M_{i}=0.678$
($\Lambda=1$),  $M_{i}=0.734$ ($\Lambda=2$), and $M_{i}=0.755$
($\Lambda=3$).  

\begin{figure}[htbp!]
  \includegraphics[width=0.49\textwidth]{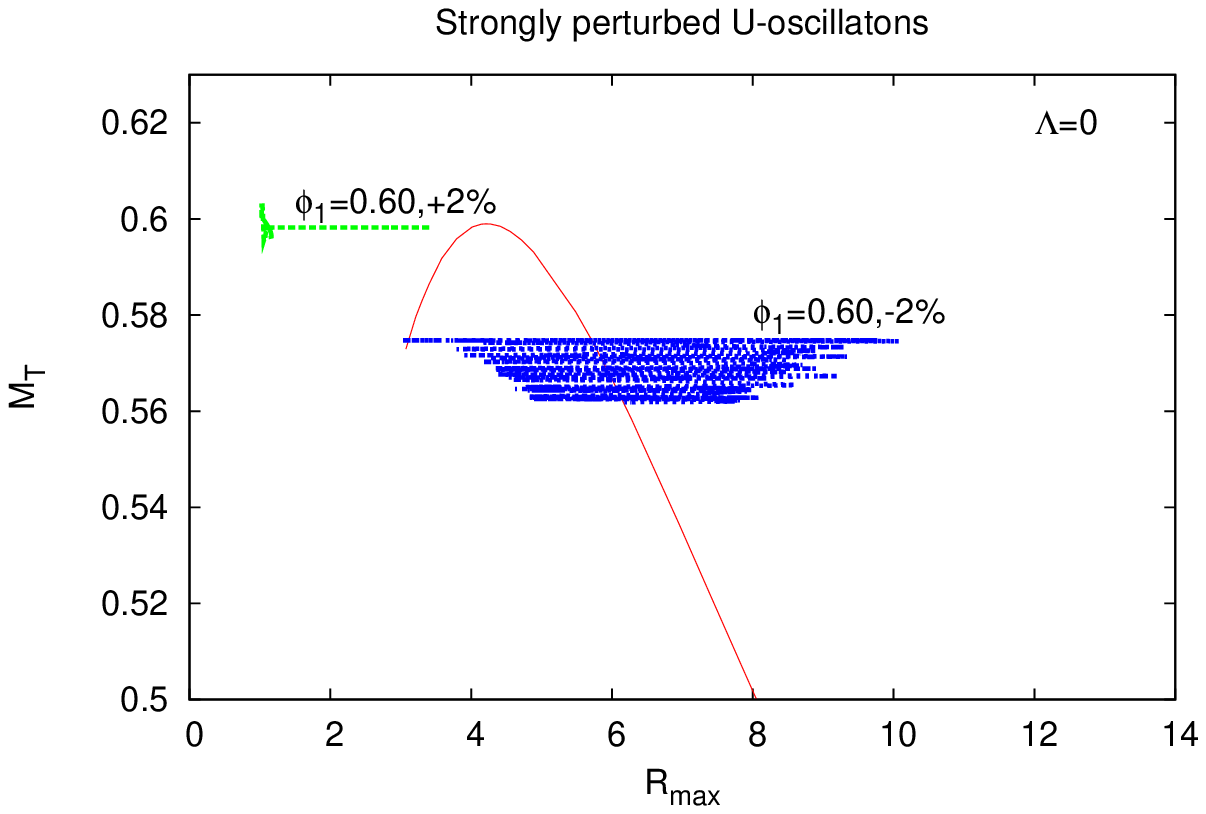}
  \includegraphics[width=0.49\textwidth]{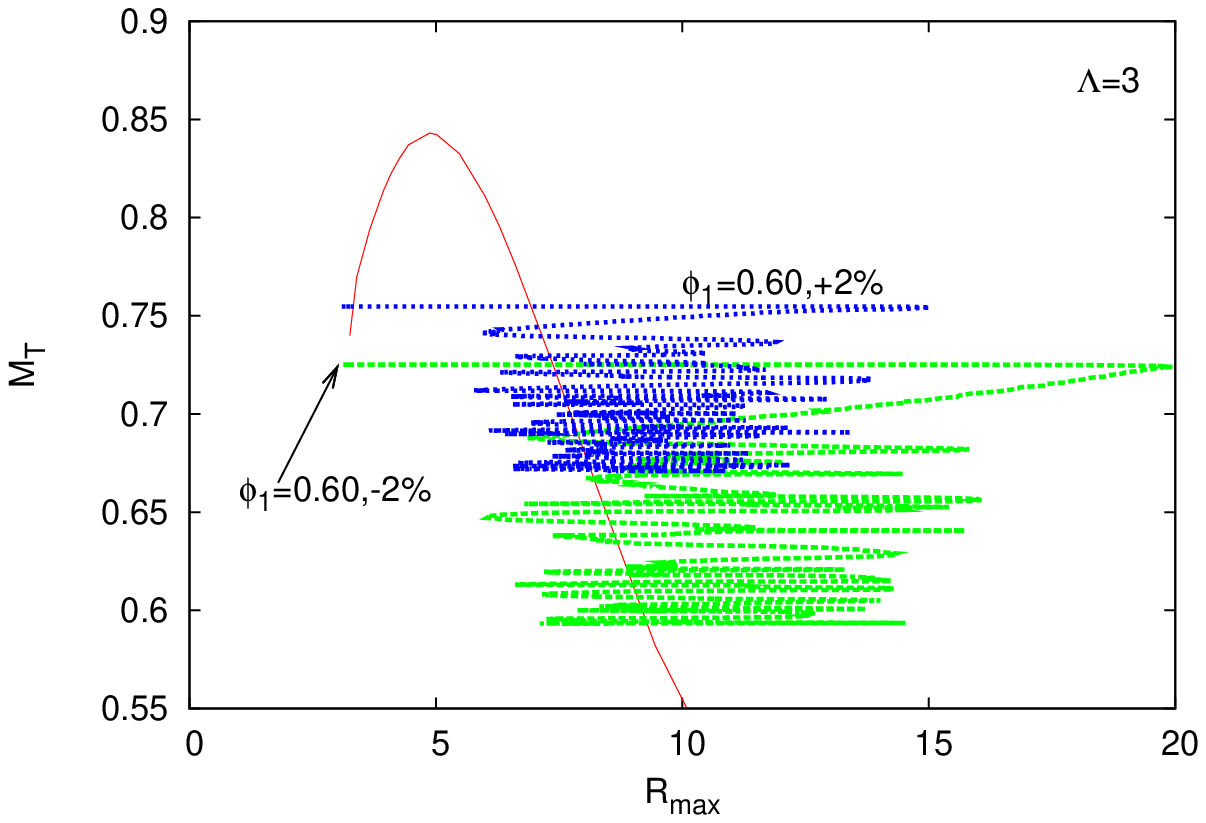}
  \caption{Evolution of the perturbed configurations with
    $\phi_{1}(0)=0.5$ and $\phi_{1}(0)=0.6$ in a plot of $M_{T}$
    versus $R_{max}$. Only in the case of $\Lambda=0$, both
    configurations collapse into a black hole if their mass is
    increased, even by a little. For other values of $\Lambda$, the
    larger values of the critical mass are able to prevent the
    collapse of the less perturbed of the configurations.} 
   \label{up}
\end{figure}

In the case of $\Lambda =0$, this configuration collapses into a
black hole, as reported also in\cite{Alcubierre:2003sx}. But, for
other $\Lambda$ values, the same configuration is able to migrate to
the S-branch, see Fig.~\ref{up}. The migration path for this
configuration is labeled as $C$, for $\Lambda=1,2,3$, in
Fig.~\ref{quasi}. In contrast, a mass increase by $5\%$ for the
equilibrium configuration with $\phi_{1}(0)=0.5$ provokes a rapid
collapse into a black hole, independently of the value of $\Lambda$.

In another experiment, we decrease the original mass of a
$\phi_{1}(0)=0.6$-oscillaton by $2\%$. The initial masses obtained
are: $M_{i}=0$.$574$ ($\Lambda=0$),  $M_{i}=0$.$651$  ($\Lambda=1$),
$M_{i}=0$.$706$ ($\Lambda=2$), and $M_{i}=0$.$725$ ($\Lambda=3$). As
expected, these perturbed U-oscillatons lose mass and migrate to the
S-branch. The evolution of these strongly perturbed configurations
appears in Fig.~\ref{up}, and their path migration is labeled as $D$
in Fig.~\ref{quasi} for all the values of $\Lambda$.

\begin{figure}[htbp!]
  \includegraphics[width=0.49\textwidth]{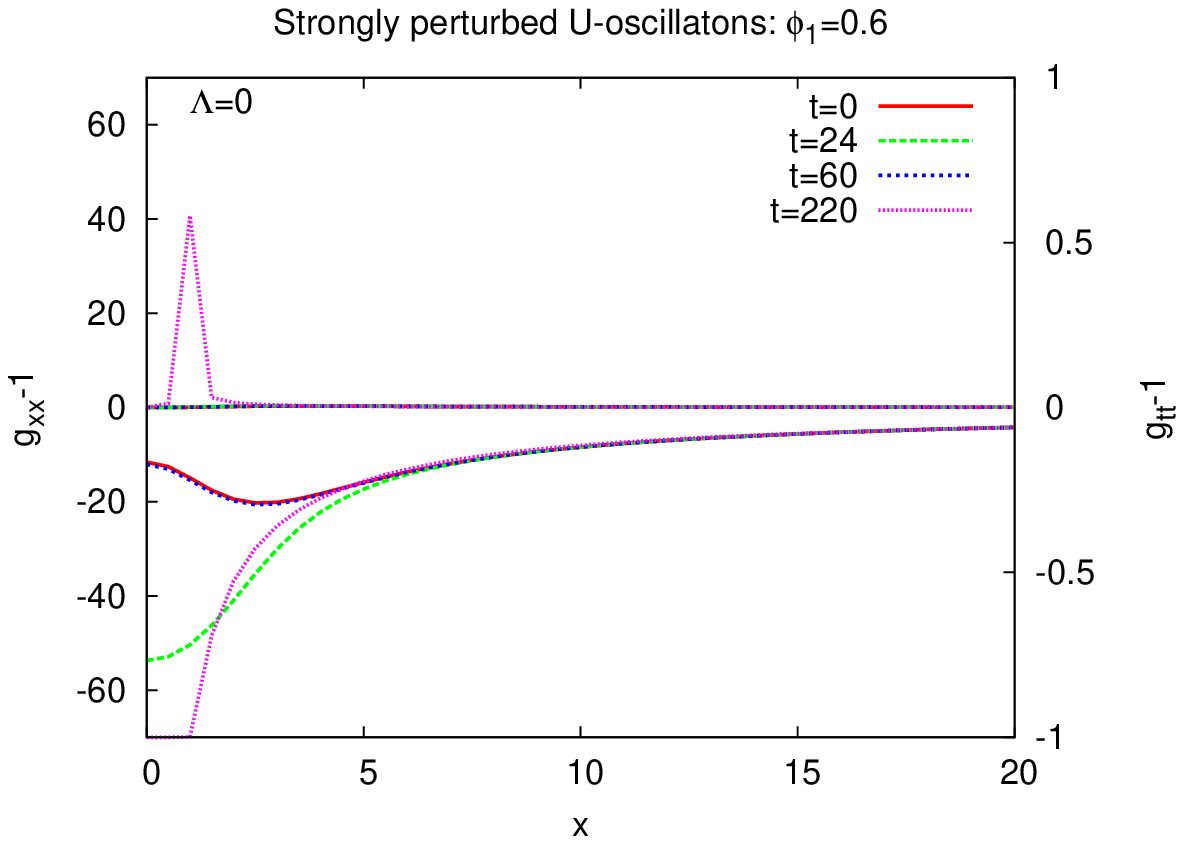}
  \includegraphics[width=0.49\textwidth]{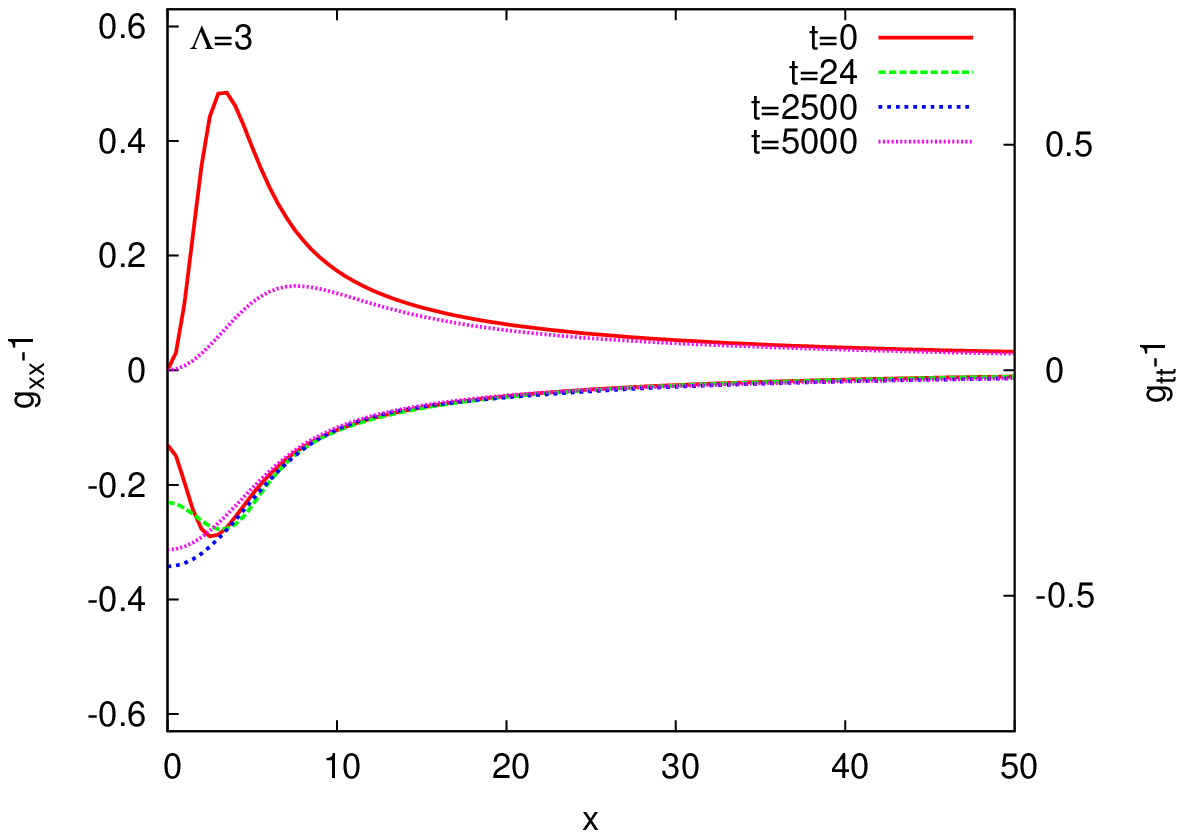}
  \caption{Radial profiles of the metric functions $g_{rr}$ and $g_{tt}$ for a
    strongly-perturbed U-configuration corresponding to
    $\phi_{1}(0)=0.6$ with an excess mass of $2\%$. The formation of a
    black hole seems to happen for $\Lambda = 0$, but its formation is
    prevented for other values.} 
   \label{metricfun}
\end{figure}

Fig.~\ref{metricfun} shows the behavior of the metric functions
$g_{rr}$ and $g_{tt}$ for the case of strongly perturbed
U-oscillaton. In the plot for $\Lambda=0$, we can see that $g_{tt}$
shows the well known "collapse of the lapse", and $g_{rr}$ shows the "grid
stretching"; both phenomena signal the possible formation of a black
hole. These behaviors are absent in the evolution of the metric
functions for other values of $\Lambda$.  

\begin{figure}[htbp!]
  \includegraphics[width=0.49\textwidth]{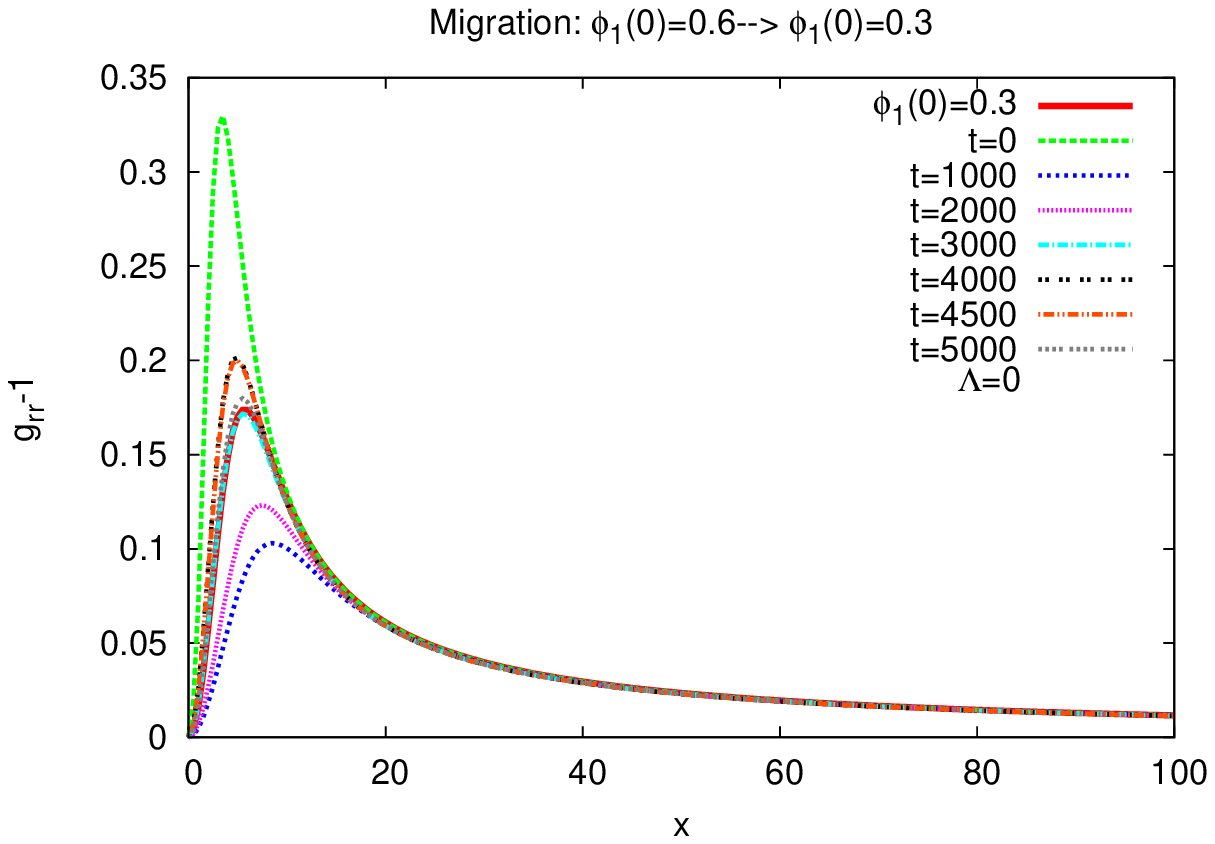}
  \includegraphics[width=0.49\textwidth]{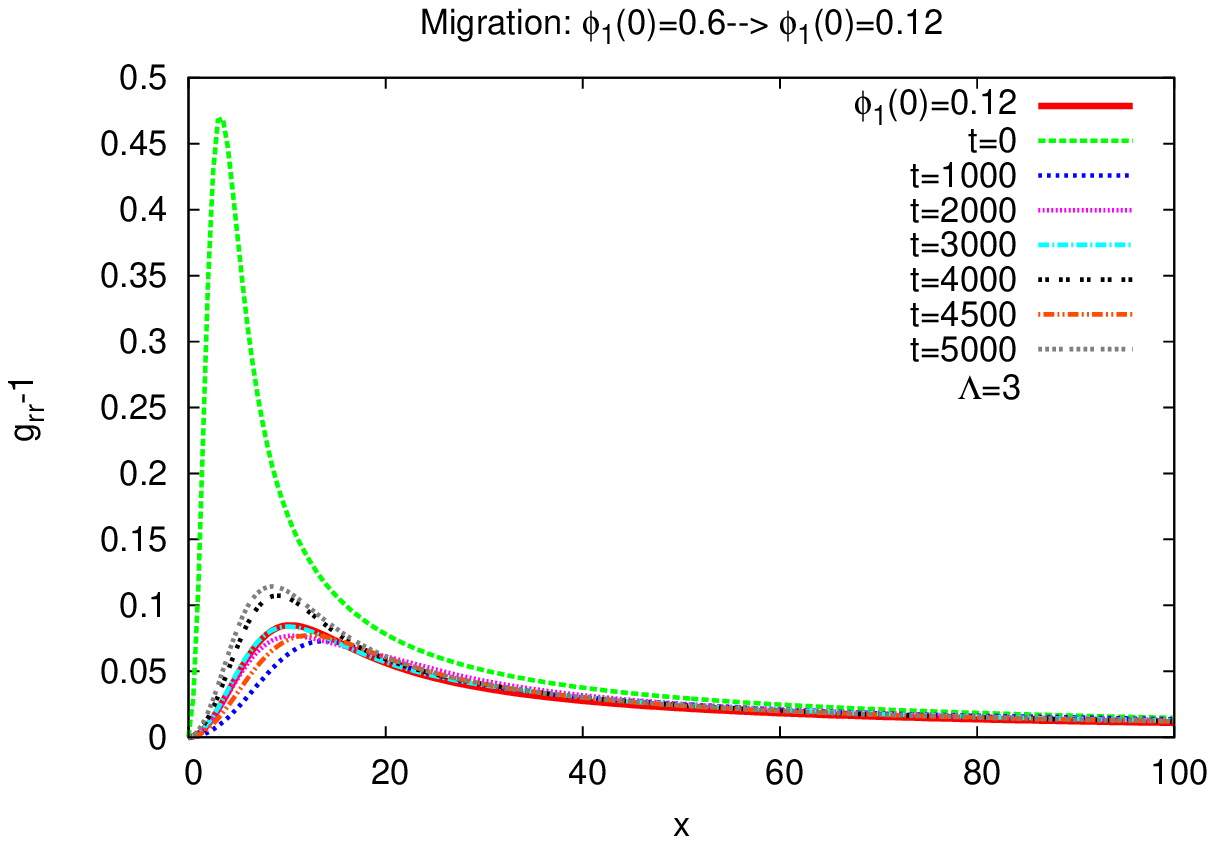}
  \caption{The evolved profiles of the radial metric function $g_{rr}$
    of a perturbed U-oscillaton with $\phi_{1}(0)=0.6$ and different
    values of $\Lambda$. In both cases, the evolution goes towards an
    equilibrium configuration in the S-branch.} 
   \label{migration}
\end{figure}

Migration of the strongly-perturbed U-oscillaton corresponding to
$\phi_{1}(0)=0.6$, with a mass decrease of $2\%$, towards a
S-oscillaton. See also Fig.~\ref{migration}, where we show the evolved
profile of the radial metric function $g_{rr}$ of the
$\phi_{1}(0)=0.6$-oscillaton migrating to the following S-branch
configurations: $\phi_{1}(0)=0.3$ for $\Lambda=0$, $\phi_{1}(0)=0.28$
for $\Lambda=1$, $\phi_{1}(0)=0.18$ for $\Lambda=2$, and
$\phi_{1}(0)=0.12$ for $\Lambda=3$.
 
\subsubsection{\label{sec:s-u-transition}The S-U transition point}

\begin{figure}[htbp!]
  \includegraphics[width=0.49\textwidth]{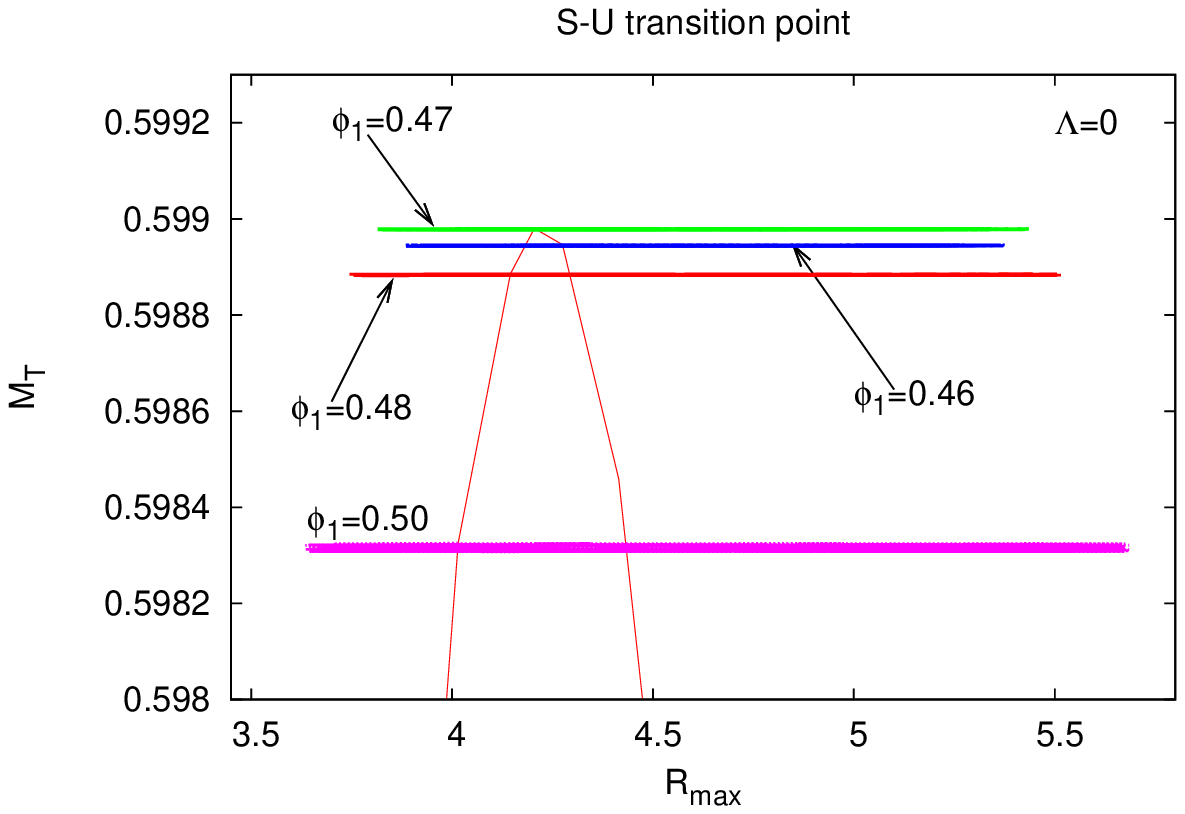}
  \includegraphics[width=0.49\textwidth]{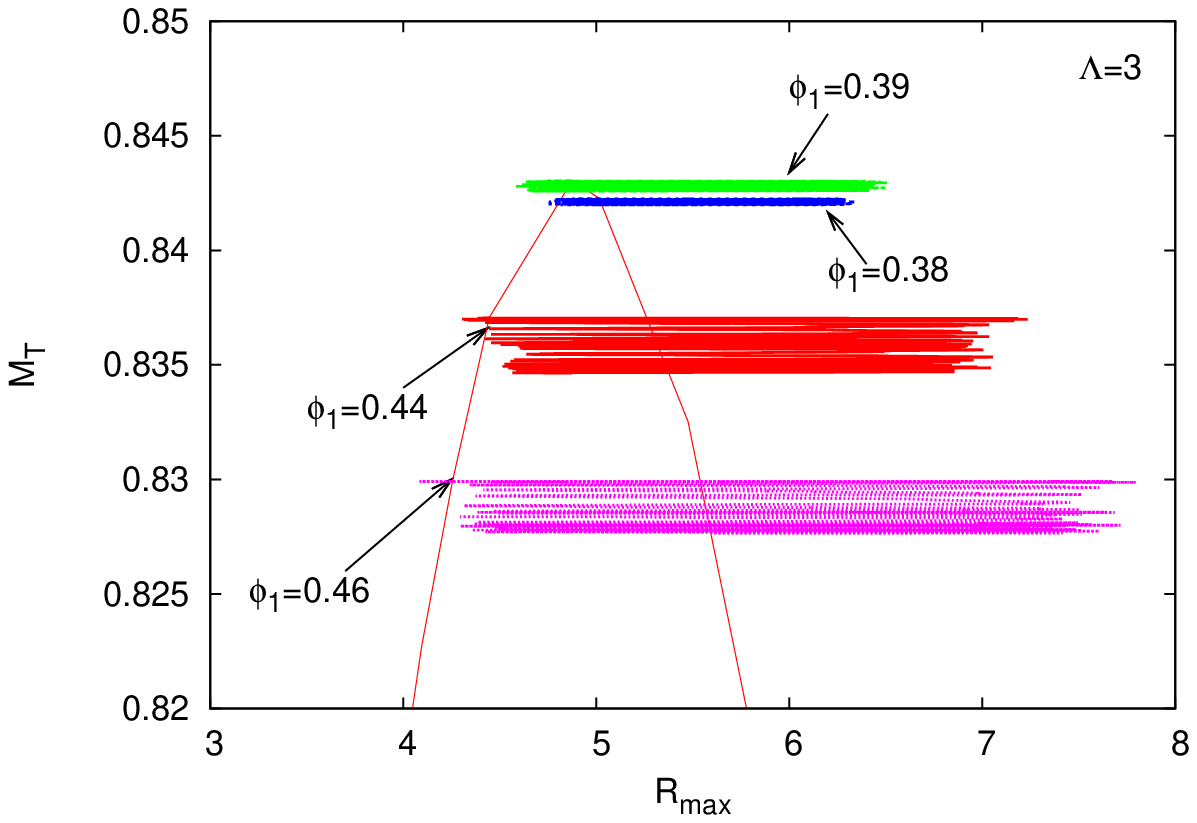}
  \caption{Evolution of the slightly perturbed configuration near the
    S-U transition point. The nature of the transition point does not
    depend upon the value of $\Lambda$.} 
   \label{transicion}
\end{figure}

We look carefully at the equilibrium configurations located nearby the
critical configuration, .i.e., the most massive equilibrium
configuration, see for instance Fig.~\ref{masas}, which is usually
called the S-U transition
point\cite{Alcubierre:2003sx}. Fig.~\ref{transicion} shows the
evolution of some slightly perturbed configuration near to the S-U
transition point. With these examples we confirm previous results:
S-oscillatons are intrinsically stable, whereas U-oscillatons are
intrinsically unstable, no matter its proximity to the S-U transition
point. Thus, this also confirms that the critical configuration is a
true stability-instability transition point.

\section{\label{sec:conclusions}Conclusions}

In this work, we solved numerically the Einstein-Klein-Gordon
system for a massive and real scalar field endowed with a scalar
potential containing a quartic self-interaction term. The diverse
numerical experiments confirm the same general properties found in the
case of the free massive case, which are also present in the case of
boson stars.

For all cases of the quartic self-interaction, it is found that there
is a critical equilibrium configuration, which is the most massive one
in each case; this critical configuration allows the separation of
configurations into stable and unstable ones, the so-called S and U
branches. As expected, the mass of the critical configuration
increases for larger values of the quartic interaction.

Equilibrium configurations located on the S-branch are intrinsically
stable, and vibrate with definite quasi-normal frequencies when
slightly perturbed. If strongly perturbed, they can either migrate to
another S-configuration, or collapse into a black hole, if its initial
mass is smaller or larger, respectively, than the critical one. On the
other hand, U-oscillatons are intrinsically unstable
configurations. Under small perturbations, they are able to lose mass
and migrate towards and equilibrium configuration on the S-branch, as
long as its initial mass is not bigger than the critical one. However,
under strong perturbations, U-oscillatons cannot prevent its collapse
into black holes.

All in all, we have found that oscillatons with a quartic
self-interaction share similar properties with their boson star
counterparts. The larger the quartic interaction, the larger values
the critical mass takes, but also the space for stable configurations
is reduced because more diluted values of the scalar field are needed
to sustain them. This can be seen in the shift of the critical scalar
field value towards zero as the quartic interaction is increased.

It proved quite difficult to obtain equilibrium configurations for
large values of the quartic interaction, because of the larger system
of equations that must be solved in the case oscillatons (in terms of
a Fourier expansion) as compared to the case of boson
stars. Nonetheless, we have been able to explore the space parameter
beyond the free case ($\Lambda = 0$), and provided evidence that
points out to the close similarity of scalar configurations, whether
we speak of real or complex scalar fields.

\acknowledgments{SV-A thanks Carlos Palenzuela for useful comments, and
  acknowledges support from CONACyT, M\'exico, and the kind
  hospitality of the Canadian Institute for Theoretical Astrophysics
  (CITA), for a short research stay during which part of this work was
  done. RB acknowledges support from CONACyT, M\'exico under grants
  83825. LAU-L thanks the Berkeley Center for Cosmological Physics
  (BCCP) for its kind hospitality, and the joint support of the
  Academia Mexicana de Ciencias and the United States-Mexico
  Foundation for Science for a summer research stay at BCCP. This work
  was partially supported by PROMEP, DAIP-UG, and by CONACyT M\'exico
  under grants 56946, and I0101/131/07 C-234/07 of the Instituto
  Avanzado de Cosmologia (IAC) collaboration.}
 
\bibliography{artref}

\end{document}